\documentclass[12pt]{article}
\pdfoutput=1
\usepackage[utf8]{inputenc}
\usepackage[T1]{fontenc}
\usepackage{amsmath,amssymb,mathtools}
\usepackage{slashed}
\usepackage{cancel}
\usepackage{graphicx}
\usepackage{booktabs,multicol,multirow}
\usepackage[shortlabels]{enumitem}
\usepackage{hyperref}
\usepackage[capitalize]{cleveref}
\usepackage{xspace}
\usepackage[noblocks]{authblk}
\usepackage[dvipsnames]{xcolor}
\usepackage{soul}
\usepackage[textsize=tiny]{todonotes}
\usepackage{subcaption}
\usepackage[normalem]{ulem}
\usepackage{nicefrac}
\usepackage{bm}
\usepackage{bbold}
\usepackage{units}
\usepackage{makecell}
\usepackage[frozencache=true]{minted}
\definecolor{bg}{rgb}{0.95,0.95,0.95}
\usepackage[sorting=none,%
  citestyle=numeric-comp,%
  bibstyle=mynumeric,%
  giveninits=true,backend=bibtex]{biblatex}

\newcommand{\MS}{{\ensuremath{\overline{\text{MS}}}}\xspace}

\newcommand{\OS}{{\ensuremath{\text{OS}}}\xspace}
\newcommand{\cp}{\ensuremath{\mathcal{CP}}\xspace}

\newcommand{\Ztwo}{\mathbb{Z}_2}

\newcommand{\nn}{\nonumber}
\newcommand{\order}[1]{\ensuremath{\mathcal{O}(#1)}\xspace}
\newcommand{\alt}{\alpha_t}

\newcommand{\tev}{\,\, \mathrm{TeV}}
\newcommand{\gev}{\,\, \mathrm{GeV}}

\newcommand{\ie}{\textit{i.e.}\xspace}
\newcommand{\eg}{\textit{e.g.}\xspace}
\newcommand{\cf}{\textit{cf.}\xspace}

\newcommand{\anyH}{\texttt{anyH3}\xspace}

\newcommand{\FeynArts}{\texttt{FeynArts}\xspace}

\newcommand{\OneCalc}{\texttt{OneCalc}\xspace}
\newcommand{\TwoCalc}{\texttt{TwoCalc}\xspace}

\newcommand{\py}{\texttt{Python}\xspace}

\newcommand{\Tint}{\texttt{Tintegrals}\xspace}

\newcommand{\olct}{\ensuremath{\delta^{(1)}_\text{CT}}}

\NewBibliographyString{refname}
\NewBibliographyString{refsname}
\DefineBibliographyStrings{english}{%
  refname = {Ref\adddot},
  refsname = {Refs\adddot}
}
\DeclareCiteCommand{\ccite}
  {%
  \ifnum\thecitetotal=1
    \bibstring{refname}%
  \else%
    \bibstring{refsname}%
  \fi%
  \addnbspace\bibopenbracket%
  \usebibmacro{cite:init}%
   \usebibmacro{prenote}}
  {\usebibmacro{citeindex}%
   \usebibmacro{cite:comp}}
  {}
  {\usebibmacro{cite:dump}%
   \usebibmacro{postnote}%
   \bibclosebracket}
\newrobustcmd*{\Ccite}{\bibsentence\ccite}

\newmintinline[code]{Mathematica}{}

\newcommand{\llog}{\overline{\ln}}
\newcommand{\zssm}{$\mathbb{Z}_2$SSM}

\newcommand{\kapSH}{\kappa_{SH}}
\newcommand{\kapS}{\kappa_{S}}
\newcommand{\lnMS}{\overline{\ln}m_s^2}
\newcommand{\lnMSs}{\overline{\ln}^2m_s^2}

\evensidemargin \oddsidemargin
\begin{document}

\thispagestyle{empty}
\def\thefootnote{\fnsymbol{footnote}}

\begin{flushright}
\texttt{DESY-24-204}\\
\texttt{FR-PHENO-2024-10}
\end{flushright}
\vspace{3em}
\begin{center}
{\Large{\bf
Generic two-loop results for trilinear\\[.7em]and quartic scalar self-interactions
}}
\\
\vspace{3em}
 {
Henning Bahl$^{1}$\footnotetext[0]{bahl@thphys.uni-heidelberg.de, 
johannes.braathen@desy.de, martin.gabelmann@physik.uni-freiburg.de},
Johannes Braathen$^{2}$,
Martin Gabelmann$^{3}$,
Sebastian Paßehr$^{4}$
 }\\[2em]
 {\sl $^1$ Institut für Theoretische Physik, Universität Heidelberg, Philosophenweg 16,\\ 61920 Heidelberg, Germany}\\[0.2em]
 {\sl $^2$ Deutsches Elektronen-Synchrotron DESY, Notkestr.~85, 22607 Hamburg, Germany}\\[0.2em]
  {\sl $^3$   Albert-Ludwigs-Universität Freiburg, Physikalisches Institut,\\
Hermann-Herder-Str.~3, 79104 Freiburg, Germany}\\[0.2em]
 {\sl $^4$ Institute for Theoretical Particle Physics and Cosmology,\\
RWTH Aachen University, Sommerfeldstraße 16, 52074 Aachen, Germany\footnote{Former address.}}
\def\thefootnote{\arabic{footnote}}
\setcounter{page}{0}
\setcounter{footnote}{0}
\end{center}
\vspace{2ex}

\begin{abstract}

Reconstructing the shape of the Higgs potential realised in Nature is a central part of the physics programme at the LHC and future colliders. In this context, accurate theoretical predictions for trilinear and quartic Higgs couplings are becoming increasingly important. In this paper, we present results that enable significant progress in the automation of these calculations at the two-loop level in a wide range of models. Specifically, we calculate the generic two-loop corrections for scalar $n$-point functions with $n\le 4$ assuming that all external scalars are identical. Working in the zero-momentum approximation, we express the results in terms of generic couplings and masses. Additionally, by exploiting permutation invariances, we reduce the number of Feynman diagrams appearing to a substantially smaller set of basis diagrams. To ease the application of our setup, we also provide routines that allow to map our generic results to scalar two-loop amplitudes generated with the package \texttt{FeynArts}. We perform a series of calculations to cross-check our results with existing results in the literature. Moreover, we present new two-loop results for the trilinear Higgs coupling in the general singlet extension of the Standard Model. We also present the public \texttt{Python} package \texttt{Tintegrals}, which allows for fast and stable evaluations of all relevant two-loop integrals with vanishing external momenta.

\end{abstract}
\setcounter{footnote}{0}
\renewcommand{\thefootnote}{\arabic{footnote}}

\newpage
\tableofcontents
\newpage

\section{Introduction}
Understanding the nature of electroweak symmetry breaking (EWSB) is one of the major goals of the Large Hadron Collider (LHC) programme, and will continue to be an essential part of physics to be explored at any future collider(s). Currently, all measurements of the properties of the discovered Higgs boson are compatible with the predictions of the Standard Model (SM) within the experimental and theoretical uncertainties, but only very little is known so far about the Higgs potential that is at the origin of EWSB. Indeed, we only know the position of the electroweak (EW) vacuum as well as the curvature of the potential at the minimum (via the measurement of the Higgs mass, see \eg~\ccite{CMS:2020xrn,ATLAS:2023owm}). In order to probe the nature of electroweak symmetry breaking, it is, however, crucial to constrain the shape of the Higgs potential also away from the EW minimum. At the LHC and possible future colliders, this is possible by measuring or constraining the trilinear and quartic Higgs couplings --- $\lambda_{hhh}$ and $\lambda_{hhhh}$, respectively --- via processes like di-Higgs production.\footnote{See also \eg~\ \ccite{Fuks:2017zkg,Papaefstathiou:2020lyp,Stylianou:2023xit,Brigljevic:2024vuv} and references therein for example studies of triple-Higgs production to constrain trilinear and quartic Higgs couplings and BSM parameter spaces.}

The existing bounds on the trilinear Higgs coupling, based primarily on searches for di-Higgs production, leave room for sizeable deviations caused by Beyond-the-SM (BSM) physics. In particular, the coupling modifier $\kappa_\lambda$, defined as
\begin{align}
\label{eq:def_kaplam}
    \kappa_\lambda\equiv \frac{\lambda_{hhh}}{(\lambda_{hhh}^\text{SM})^{(0)}}\,,
\end{align}
where $(\lambda_{hhh}^\text{SM})^{(0)}$ denotes the tree-level SM prediction for the trilinear Higgs coupling, is bound to be in the range $-1.2<\kappa_\lambda<7.2$ at the 95\% confidence level (C.L.) by ATLAS searches for di-Higgs production~\cite{ATLAS:2024ish} and assuming all other couplings to be SM-like (see also Refs.~\cite{ATLAS:2022kbf,CMS:2022dwd}). Limits on $\kappa_\lambda$ including also data from single-Higgs production --- allowing all other relevant couplings to float --- have also been obtained by both ATLAS: $-1.4 < \kappa_\lambda < 6.1$~\cite{ATLAS:2022jtk}; and CMS: $-1.4 < \kappa_\lambda < 7.8$~\cite{CMS:2024awa}.\footnote{Note that the constraints including single-Higgs data currently do not consider the theoretical uncertainty stemming from different treatments of the top-quark mass, see \eg~the discussions in~\ccite{Baglio:2020wgt,Jaskiewicz:2024xkd}.} The quartic Higgs coupling is much less constrained --- indeed, currently only perturbative unitarity provides loose bounds on the quartic Higgs coupling modifier (see \eg \ccite{Stylianou:2023xit}). Intriguingly, these seemingly weak constraints are already powerful enough to exclude so far unconstrained parameter space in multi-Higgs models~\cite{Bahl:2022jnx,Bahl:2023eau}. The experimental constraints on $\kappa_\lambda$ are projected to significantly tighten during future LHC runs and at future colliders (see \ccite{deBlas:2019rxi} for a review). As a concrete example, the projection for the high-luminosity phase of the LHC (HL-LHC) for $\kappa_\lambda$ at 68\% C.L.\ is $0.74<\kappa_\lambda<1.29$~\cite{CMS:2025hfp}, while further improvements are expected at future lepton colliders (see e.g.\ Refs.~\cite{LinearColliderVision:2025hlt,Maura:2025rcv}. Additionally, first constraints on the quartic Higgs coupling, beyond those from perturbative unitarity, are expected to be obtained at the HL-LHC~\cite{Stylianou:2023xit}. 

Given these experimental prospects, accurate theoretical predictions of $\kappa_\lambda$ become mandatory. The one-loop calculation of $\kappa_\lambda$ can be considered as solved. Indeed, many results for specific models have been derived in the literature~\cite{Barger:1991ed,Hollik:2001px,Dobado:2002jz,Williams:2007dc,Williams:2011bu,Kanemura:2015fra,Kanemura:2016lkz,He:2016sqr,Kanemura:2017wtm,Kanemura:2002vm,Kanemura:2004mg,Kanemura:2015mxa,Arhrib:2015hoa,Kanemura:2016sos,Kanemura:2017wtm,Falaki:2023tyd,Aoki:2012jj,Chiang:2018xpl,Bahl:2022gqg,Kanemura:2017gbi,Kanemura:2019slf,Aiko:2023xui,Basler:2018cwe,Basler:2020nrq,Basler:2024aaf,Nhung:2013lpa,Cherchiglia:2024abx}, and, moreover, the multi-purpose tool \texttt{anyH3}, aimed at the calculation of $\kappa_\lambda$ in all renormalisable models, has been presented in~\ccite{Bahl:2023eau}. In contrast, two-loop calculations have so far only been performed in a small number of BSM models featuring extended Higgs sectors~\cite{Senaha:2018xek,Braathen:2019pxr,Braathen:2019zoh,Braathen:2020vwo,Aiko:2023nqj} as well as supersymmetric (SUSY) models~\cite{Brucherseifer:2013qva, Muhlleitner:2015dua,Borschensky:2022pfc}. Studies in non-SUSY models have found that the size of the two-loop corrections can reach up to 70\% of the size of the one-loop corrections~\cite{Bahl:2022jnx} and their inclusion is therefore crucial to obtain reliable results for $\kappa_\lambda$. Meanwhile, in SUSY models, two-loop contributions to $\kappa_\lambda$ can considerably reduce the theoretical uncertainty for di-Higgs production~\cite{Borschensky:2022pfc}. Consequently, there is strong motivation to include two-loop corrections in calculations of $\kappa_\lambda$, in order to obtain accurate theoretical predictions. Similar conclusions have also been reached for the quartic Higgs coupling~\cite{Braathen:2019pxr,Braathen:2019zoh}.

The present work takes a big step towards an automatised two-loop prediction of the trilinear and quartic Higgs couplings in general renormalisable theories. In the spirit of earlier works for tadpoles and self-energies in the literature --- in particular \ccite{Goodsell:2019zfs}, and also Refs.~\cite{Martin:2001vx,Martin:2003it,Martin:2003qz,Martin:2018emo,Goodsell:2015ira,Staub:2013tta,Staub:2015kfa} ---, we present novel generic two-loop results for scalar three- and four-point functions (with identical external states), together with results for one- and two-loop functions in corresponding conventions. These constitute the core ingredients for two-loop predictions of $\lambda_{hhh}$ (or equivalently $\kappa_\lambda$) and of $\lambda_{hhhh}$ in arbitrary renormalisable theories. In addition, the scalar four-point function can also be used for deriving matching conditions in effective field theory calculations, see for instance \ccite{Bagnaschi:2017xid,Athron:2017fvs,Bahl:2018jom,Braathen:2018htl,Gabelmann:2018axh,Bahl:2020jaq,Slavich:2020zjv,Bagnaschi:2019esc}.

In this work, we employ the effective-potential, or in other words the zero-momentum, approximation of evaluating all amplitudes with vanishing external momenta. The motivation for this is threefold: \textit{(i)} the coupling modifiers for the trilinear and quartic Higgs couplings that are (or will be) constrained by experiments are defined for effective couplings, without including any dependence on external momenta; \textit{(ii)} it was found in \ccite{Bahl:2023eau} that the impact of momentum-dependent effects is moderate, especially for scenarios with significant BSM contributions; and lastly \textit{(iii)} two-loop three- and four-point integrals with general external momenta are typically not known for generic internal masses (with only few exceptions), and, while calculations with numerical methods~\cite{Binoth:2000ps,Binoth:2003ak,Bogner:2007cr,Heinrich:2008si,Borowka:2017idc,Moriello:2019yhu,Borinsky:2020rqs,Smirnov:2021rhf,Liu:2022chg,Borinsky:2023jdv,Heinrich:2023til} are in principle possible, these are computationally very intensive (and thus impractical for phenomenological scans of BSM models). Besides providing results for genuine two-loop vertex functions, we also derive the corresponding two-loop subloop renormalisation contributions which allow us to easily implement different renormalisation schemes. To ease the application of our results, we also provide example routines to map our generic results to a specific model with the help of the package \texttt{FeynArts}~\cite{Kublbeck:1990xc,Hahn:2000kx}.

Our paper is structured as follows. In \cref{sec:generic_calc}, we present the generic calculation of the scalar vertex functions. \cref{sec:FA_mapping} discusses the mapping to concrete models. In \cref{sec:cross-checks}, we present various cross-checks of our generic result. Novel two-loop results for the general scalar singlet extension of the SM are shown in \cref{sec:applications}. We draw conclusions in \cref{sec:conclusions}.


\clearpage

\section{Generic calculation}
\label{sec:generic_calc}

In this Section, we describe the calculation of our generic results.


\subsection{Calculational setup}
\label{sec:setup}

We begin by generating the amplitudes for the generic diagrams using \FeynArts~\cite{Kublbeck:1990xc,Hahn:2000kx} at the class level. The two-loop amplitudes are then processed using \TwoCalc~\cite{Weiglein:1993hd,Weiglein:1995qs}, while for the diagrams with one-loop counterterm insertions, we employ \OneCalc~\cite{Weiglein:1993hd,Weiglein:1995qs}.


\subsection{Canonical form of the diagrams}
\label{sec:canonicaledges}

\begin{figure}
    \centering
    \includegraphics[width=.7\textwidth]{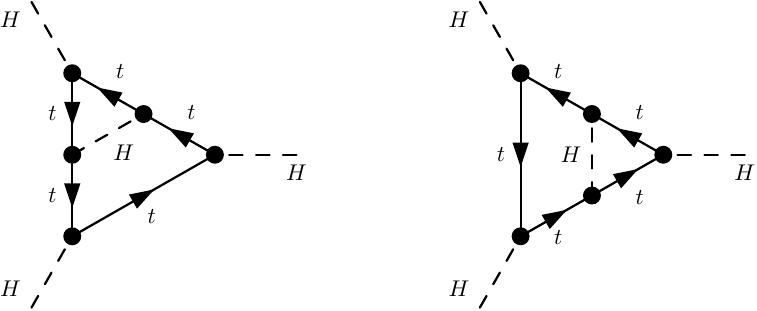}
    \caption{Example diagrams of \order{\alt^2} contributing to the trilinear Higgs coupling in the SM.}
    \label{fig:HHH_atat_example}
\end{figure}

Following \ccite{Goodsell:2019zfs}, we represent our Feynman diagrams in terms of an \emph{edge list}.  For instance, the representation of the Feynman diagrams in \cref{fig:HHH_atat_example} in the form of edge lists is as follows
\begin{minted}[bgcolor=bg]{Mathematica}
edgelist1 = {edge[v[1], v[4],  S[1]], edge[v[2], v[5],  S[1]], 
             edge[v[3], v[6],  S[1]], edge[v[4], v[7], -F[3]], 
             edge[v[4], v[8],  F[3]], edge[v[5], v[6],  F[3]], 
             edge[v[5], v[8], -F[3]], edge[v[6], v[7],  F[3]], 
             edge[v[7], v[8],  S[1]]}
edgelist2 = {edge[v[1], v[4],  S[1]], edge[v[2], v[5],  S[1]], 
             edge[v[3], v[6],  S[1]], edge[v[4], v[5],  F[3]], 
             edge[v[4], v[7], -F[3]], edge[v[5], v[8],  F[3]], 
             edge[v[6], v[7],  F[3]], edge[v[6], v[8], -F[3]], 
             edge[v[7], v[8],  S[1]]}
\end{minted}
where the \code{edgelist1} corresponds to the left diagram in \cref{fig:HHH_atat_example}; and \code{edgelist2} to the right diagram.  Here, each \code{edge} represents one edge (or line) in the Feynman diagram. The first argument denotes the initial vertex of the edge; the second argument, the final vertex of the edge; and, the third argument is the field which propagates along the edge. The numbering of the vertices is per se arbitrary, but, as in \ccite{Goodsell:2019zfs}, we choose to number the vertices consecutively. 

The form of the edge list derived from the Feynman diagram, however, depends on the output of \FeynArts. As a consequence, diagrams which are actually identical can be represented by different edge lists. This obviously happens for the two diagrams shown in \cref{fig:HHH_atat_example}, which are related by a simple rotation.

We resolve this ambiguity by deriving a canonical representation for each Feynman diagram. This canonical representation is designed such that edge lists for identical Feynman diagrams are indeed identical. Using a simplified pseudo-code form, the algorithm can be written in the following form:
\begin{minted}[bgcolor=bg]{text}
canonicaledges algorithm in pseudo code:
    - identify internal indices
    - identify external indices
    - generate permutations of external indices
    - generate permutations of internal indices
    - combine permutations of internal and external indices
    - permute the edge list following combined list of permutations 
    - sort list of permuted edge lists
    - return first edge list after sorting
\end{minted}
In this algorithm, a list of all possible representations of the Feynman diagram is derived before being sorted. By picking the first edge list after sorting, we ensure that the same edge list is always returned for all identical Feynman diagrams.
 
A similar algorithm has been used in \ccite{Goodsell:2019zfs}. However, in comparison to the implementation therein, we split up the permutation of the external\footnote{We note that the permutation of the external indices is only possible for identical external particles. } and internal indices, which drastically improves the speed of the algorithm (by significantly reducing memory usage). Without this change, the handling of four-point amplitudes would be significantly more difficult. 

For the calculation of $n$-point functions in concrete models, we use the \code{canonicaledges} algorithm in two ways. First, we use it to reduce the number of generic amplitudes to calculate. For example, only one of the topologies shown in \cref{fig:HHH_atat_example} needs to be calculated (and saved) in generic form (see \cref{sec:diagram_reduction}). Second, we use it to match Feynman amplitudes generated with \FeynArts to our generic results (see \cref{sec:FA_mapping}).


\subsection{Reducing the number of generic diagrams via symmetry relations}
\label{sec:diagram_reduction}

\begin{figure}
    \centering

     \begin{subfigure}[b]{0.3\textwidth}
         \centering
         \includegraphics[height=4.15cm]{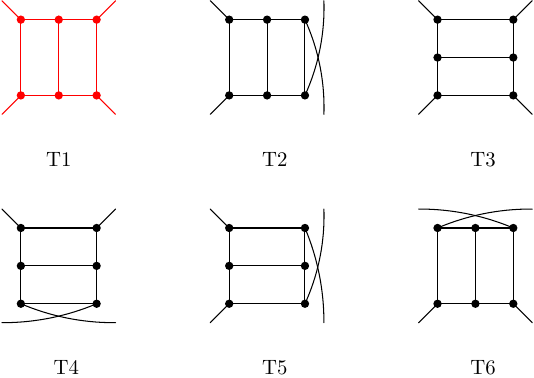}
         \label{fig:topreduction}
     \end{subfigure}
     \hspace{3cm}{
     \begin{subfigure}[b]{0.3\textwidth}
         \centering
         \includegraphics[height=5cm]{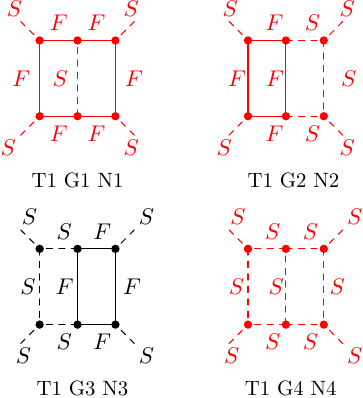}
        \label{fig:diagreduction}
    \end{subfigure}}
    \caption{The reduction of the number of graphs at the topology level and the level of generic Feynman diagrams at the example of the double-box topology. For this particular example, diagrams with internal vector bosons are not shown. Topologies/diagrams in black are taken into account by applying symmetry factors onto the appropriate red diagrams rather than being computed explicitly (\ie\ 3 instead of 24 generic diagrams need to be computed in total).}
    \label{fig:reduction}
\end{figure}

One of the main motivations for the results derived in this work is their application to precision calculations of the effective trilinear and quartic self-couplings of the discovered Higgs boson. In this case, one only needs to compute diagrams with identical external legs, which introduces many additional symmetries between diagrams that would otherwise not have been identical (for different external particles). Such symmetries can be found by considering the permutations of internal and external fields, which is at the heart of the canonical edges algorithm described in the previous section. Since our algorithm provides a unique edge list, two diagrams are identical if their edge lists are identical. We exploit this feature to reduce both the number of topologies to consider and the number of generic diagrams to compute. To construct a minimal and complete list of Feynman diagrams to be computed, we start by generating all possible one- and two-loop topologies using \FeynArts, then calculate the corresponding edge lists, and finally remove identical topologies while simultaneously increasing a symmetry factor for them. Once a minimal set of topologies is found, we use \FeynArts to insert all possible generic fields of spin 0, \nicefrac{1}{2}, or 1 for a renormalisable QFT. The resulting list of Feynman diagrams is again reduced by comparing edge lists and introducing appropriate symmetry factors.

In the left panel of \cref{fig:reduction}, we demonstrate the reduction of topologies and diagrams for the example of the four-point double-box diagram. The topologies \texttt{T1-T6} are in general different if the external states and momenta are not identical. However, for the calculation of the effective trilinear self-coupling, all external lines are the same and the computed edge lists are identical. Therefore, we only consider \texttt{T1} and introduce an additional symmetry factor of 6. In the next step, we fix all external lines to exactly the same generic scalar field and populate the internal lines with generic states of all possible spin. In the right panel of \cref{fig:reduction}, we restricted the algorithm to only consider spin 0 (\texttt{S}) and spin \nicefrac{1}{2} (\texttt{F}) states for demonstrative purposes as the inclusion of vectors would greatly increase the number of diagrams to draw. In this particular sub-class of diagrams, there exists merely one diagram that contains only scalar fields. We note that this (and all other) diagram(s) inherits the symmetry factor 6 from the reduction of the topologies. In addition, \FeynArts' \textit{InsertFields} function generates three more diagrams containing scalars and fermions out of which two, \texttt{T1 G2 N2} and \texttt{T1 G3 N3}, have the same edge list and are therefore identical. In total, this sub-class of diagrams contains only 3 unique graphs (coloured in red) that need to be computed compared to the original $6\times 4=24$ diagrams. 

\begin{table}[tb]
    \centering
    \begin{tabular}{c|c|c}
     $n$ &   topology-level & field-level \\ \hline
      0 & $2  \to 2$  & $11    \to 11$ \\
      1 & $3  \to 3$  & $25    \to 25$ \\
      2 & $9  \to8$   & $121   \to 92\, (102)$\\
      3 & $40 \to 13$ & $936   \to 229\, (291)$ \\
      4 & $265\to 29$ & $10496 \to 698\, (928)$
    \end{tabular}
    \caption{Resulting reduction of the number of generic two-loop topologies and generic two-loop Feynman diagrams using symmetry-relations described in the text. $n$ represents the number of external legs (scalars at the field level). Numbers in brackets are obtained if diagrams, which are symmetric to diagrams with all fields replaced by their anti-fields, are kept.}
    \label{tab:symmetryfacs}
\end{table}

In \cref{tab:symmetryfacs}, we summarise the reduction of the number of graphs at the level of topologies and at the level of generic fields (if scalars, fermions and vectors are allowed on the internal lines) at two loops for all scalar $n=0,1,2,3,4$-point functions.\footnote{By ``$0$-point functions'', we refer to vacuum bubbles, which contribute to the effective potential at the two-loop level.} While the number of vacuum, tadpole and self-energy diagrams is only slightly reduced, the presented algorithm reduces the number of three- and four-point graphs by about one order of magnitude.
It is important to note that this reduction is applied \textit{before} even considering a concrete model and that, due to specific coupling structures, some models \textit{(i)} may involve many more diagrams to be mapped onto the generic graphs and \textit{(ii)} may also introduce more internal symmetries between certain diagrams. For both reasons, the reduction of the number of diagrams is not only a crucial step to be done \textit{before} performing the tensor reduction using \TwoCalc and the integral reduction discussed in \cref{sec:integral_reduction}, but also afterwards for a given set of Feynman diagrams generated by \eg \FeynArts for a concrete model, \cf \cref{sec:FA_mapping}.


\subsection{Integral reduction}
\label{sec:integral_reduction}

All amplitudes are evaluated for zero external momentum. In this approximation, all scalar loop integrals can be written in the form of a $T$ or a $Y$ integral. Following the notation of \ccite{Weiglein:1993hd,Goodsell:2019zfs}, these are defined via
\begin{align}
\label{eq:Tint}
   T_{i_1\cdots i_n} &= \int \frac{\mathrm{d}^dq_1\,\mathrm{d}^dq_2}
   {\left[i\,\pi^2 \left(2\,\pi\,\mu\right)^{d - 4}\right]^2}\,
   \frac{1}{\left(k_{i_1}^2 - m_{i_1}^2\right) \cdots
   \left(k_{i_n}^2 - m_{i_n}^2\right)}\,,\\
   Y_{i_1\cdots i_n}^{j_1\cdots j_o}{} &=
   \int \frac{\mathrm{d}^dq_1\,\mathrm{d}^dq_2}
   {\left[i\,\pi^2 \left(2\,\pi\,\mu\right)^{d - 4}\right]^2}\,
   \frac{k_{j_1}^2 \cdots k_{j_o}^2}{\left(k_{i_1}^2 - m_{i_1}^2\right) \cdots
   \left(k_{i_n}^2 - m_{i_n}^2\right)}\,,
\end{align}
where $i_1,\dots, i_n, j_1,\dots, j_o \in \{1, 2, 3,
4, 5\}$. $d = 4 -2\epsilon$ with $\epsilon$ being the ultra-violet regulator, $q_1$ and $q_2$ are the loop momenta, and $\mu$ is the regularisation scale. The kinematic variables are given by
\begin{align}
  k_1 = q_1\,,\quad k_2 = q_1 + p\,,\quad k_3 = q_2 - q_1\,,\quad
  k_4 = q_2\,,\quad k_5 = q_2 + p,
\end{align}
where $p$ is the external momentum. We note, however, that for the rest of this work, the external momentum in $T$ or $Y$ integrals is always set to zero.

All $Y$ integrals can themselves be straightforwardly expressed in terms of $T$ integrals. These can in principle be reduced to the $T_{134}$ integral and (products of) one-loop integrals. This can, however, lead to issues during the numerical evaluation of the integrals.

Consider for example the loop integral $T_{113}(m_1,m_2,m_3)$. If all masses are different, the integral can be reduced via
\begin{align}
T_{113}(m_1,m_2,m_3) = \frac{1}{m_1^2 - m_2^2} A_0(m_3^2) \left[A_0(m_1^2) - A_0(m_2^2)\right] \,.
\end{align}
This relation poses a challenge during numerical evaluation if $m_1\to m_2$. If both masses are equal, the alternative reduction formula
\begin{align}
T_{113}(m_1,m_1,m_3) = \frac{D-2}{2 m_1^2} A_0(m_1^2) A_0(m_3^2)
\end{align}
should be used. If instead $m_2\to m_3$, the first relation can be used without problems.

To capture all these cases, we have generated a list of all possible mass configurations (with one or more masses being zero or equal to each other). To reduce the size of this list, we sort the mass arguments for each propagator type (\eg, $T_{113}(m_2,m_1,m_3) = T_{113}(m_1,m_2,m_3)$). We then derived the reduction rules for all these mass configurations, using \TwoCalc. After generating this list, we checked which rules actually need to be used for the reduction and, on the other hand, which cases are already covered by a more general rule without causing any numerical instability (as in the case of $m_2\to m_3$ for the example above) and can be discarded. In this way, the number of reduction rules is greatly reduced. In the end, we obtain a list of reduction rules and an associated list of mass configurations including information about which reduction rule should be applied.

This result can then be used to implement numerical evaluation routines that automatically choose the right reduction rule based on the numerical values used as inputs for the masses. This ``on-the-fly'' reduction allows for stable scans of parameter spaces. A similar procedure is already used \eg~in \texttt{FeynHiggs}~\cite{Heinemeyer:1998yj,Heinemeyer:1998np,Hahn:2009zz,Degrassi:2002fi,Frank:2006yh,Hahn:2013ria,Bahl:2016brp,Bahl:2017aev,Bahl:2018qog}.

To facilitate the application of our results, we have developed the new \texttt{Python} package \texttt{Tintegrals}, which is available at
\begin{center}
  \url{https://gitlab.com/anybsm/tint}
\end{center}
or via ``\mintinline{latex}{pip install Tintegrals}''. This package provides numerical routines to evaluate all the $T$ integrals appearing in our calculation using the on-the-fly reduction explained above. Further details can be found in \cref{app:Tint}.


\subsection{Structure of the generic results}

The steps discussed in \cref{sec:setup,sec:canonicaledges,sec:diagram_reduction,sec:integral_reduction}
are performed for a general renormalisable theory and therefore only need to be performed once.
We provide the results in the \texttt{Wolfram Language} format structured as follows
\begin{minted}[bgcolor=bg]{text}
{
  {
    {1, 1, 1}, 
    {
     edge[iv[1], v[5], S[i3]], edge[iv[2], v[6], S[i4]],
     edge[iv[3], v[7], S[i1]], edge[iv[4], v[8], S[i2]], 
     edge[v[5], v[6], -S[i9]], edge[v[5], v[7], -S[i6]], 
     edge[v[5], v[8], -S[i8]], edge[v[6], v[7], -S[i5]],
     edge[v[6], v[8], -S[i7]]
    },
    -((lfac^2*SSS[i1, -i5, -i6, 1]*SSS[i2, -i7, -i8, 1]*
        SSSS[i3, i6, i8, i9, 1]*SSSS[i4, i5, i7, -i9, 1]*
        T[Df[k1, MS[i7]]*Df[k1, MS[i8]]*Df[k3, MS[i5]]*
          Df[k3, MS[i6]]*Df[k4, MS[i9]]]
       )/SF[1, 1/6, 1])
  },
...
}
\end{minted}
where the first entry is the diagram identifier used by \FeynArts, the second entry is the unique edge list and the third entry is the result for the generic diagram. Therein, \texttt{SF} represents the various symmetry-factors assigned by \FeynArts, by the reductions of the topologies, and by the reduction of diagrams discussed in \cref{sec:diagram_reduction}. For evaluations in concrete models, the replacement \texttt{SF->Times} should be applied. The scalar integrals
\begin{equation}
    \texttt{T[Df[k1, m1]*...*Df[kn, mn]]} \equiv T_{i_1\dots i_n}
\end{equation}
are defined in \cref{sec:integral_reduction} and \texttt{lfac=(4*Pi)\textasciicircum(-2)} is the loop factor. The generic couplings (here \texttt{SSS[]} and \texttt{SSSS[]} for scalar triple and quartic couplings, respectively) are defined in \ccite{Goodsell:2019zfs}.

For a given set of diagrams in a given model, one needs to compute the edge list of every diagram, extract the couplings and masses of the diagram and insert them into the appropriate generic result above. This is discussed in the next section. For instance, the example shown above corresponds to the generic scalar double-box diagram shown in the lower right of \cref{fig:reduction} (denoted \texttt{T1 G4 N4}). The full lists containing all diagrams counted in \cref{tab:symmetryfacs} can be downloaded at
\begin{center}
    \url{https://gitlab.com/anybsm/generic_scalar_2l_amplitudes}.
\end{center}


\section{Mapping to \FeynArts output}
\label{sec:FA_mapping}

To apply our generic results to a specific model, a non-trivial mapping of this model onto the conventions of our generic expressions must be performed. First, all diagrams appearing in the specific model have to be identified, and then the couplings and masses of the specific model have to be mapped to their generic counterparts. While these steps are straightforward when handling a single or a few diagrams, they can quickly become cumbersome if a larger number of diagrams is involved.

We automatise these steps by providing routines to directly map \FeynArts output onto our generic results. This mapping consists of the following steps:
\begin{itemize}
    \item Use \FeynArts routines to generate the Feynman diagrams.
    \item Map these diagrams to the canonical-form diagrams used for our generic results (applying the \texttt{canonicaledges} algorithm, see \cref{sec:canonicaledges}).
    \item Identify the contributing fields.
    \item Extract the couplings that appear in the diagrams and map them to the generic couplings appearing in our generic results.
    \item Write out the generic results with the generic couplings and masses replaced by the model-specific expressions.
\end{itemize}
These steps can be executed with a single command. In this way, analytic results for $n=0,1,2,3,4$-point functions can be derived with minimal user intervention.

An exemplary implementation of this mapping can be found at
\begin{center}
    \url{https://gitlab.com/anybsm/generic_scalar_2l_amplitudes}.
\end{center}


\section{Cross-checks}
\label{sec:cross-checks}

In this Section, we cross-check our generic results by using them to derive a series of known two-loop results from the literature. Throughout the following sections, we denote the scalar $m$-point function, computed for the external fields $s_1,\dots,s_m$ at the $n$-loop order, as $\Gamma_{s_1\dots s_m}^{(n)}(p_1^2,\dots,p_m^2)$. The superscripts ``gen.'' and ``subloop'' refer to the genuine two-loop diagrams and to the two-loop subloop renormalisation contributions, \ie one-loop diagrams with one-loop counterterm insertions, respectively. The $n$-loop counterterm for a parameter $X$ is denoted by $\delta^{(n)}_{\rm CT}X$, while $n$-loop shifts to couplings $\lambda_{s_1\dots s_m}$ are denoted by $\delta^{(n)}\lambda_{s_1\dots s_m}$. Finally, we denote fully-renormalised loop-level quantities including external-leg corrections with a hat --- \eg~$\hat\lambda_{hhh}$ for the complete result for the trilinear Higgs coupling.


\subsection{Cross-checks in the Standard Model}
\label{sec:SM}

As a first verification of our results in the setting of a concrete model, we reproduce in this Section the leading $\mathcal{O}(\alpha_s\alpha_t)$ and $\mathcal{O}(\alpha_t^2)$ corrections to $\lambda_{hhh}$ and $\lambda_{hhhh}$ in the SM in the limit of neglecting effects from the SM-like Higgs boson mass in the two-loop diagrams, which were derived in Refs.~\cite{Braathen:2019pxr,Braathen:2019zoh,Senaha:2018xek}. 

We start by generating the genuine two-loop diagrams contributing at these orders to the one-, two-, three-, and four-point functions with external Higgs bosons using \texttt{FeynArts} and its in-built SM model file with QCD extension. At the level of the \FeynArts output, we obtain for the scalar three-point (four-point) function 12 (60) diagrams at $\mathcal{O}(\alpha_s\alpha_t)$ and 39 (204) at $\mathcal{O}(\alpha_t^2)$; after our reduction algorithm is applied, these numbers are reduced to 4 (6) at $\mathcal{O}(\alpha_s\alpha_t)$ and 13 (26) at $\mathcal{O}(\alpha_t^2)$. Applying then our routines to map our generic results to these diagrams, we obtain at $\mathcal{O}(\alpha_s\alpha_t)$ 
\begin{align}
    \Gamma^{(2)}_h\big|^\text{gen.}_{\mathcal{O}(\alpha_s\alpha_t)}&=\frac{\alpha_s m_t^4}{\pi^3v}\bigg[-\frac{3}{2\epsilon^2} + \frac{3\, \llog m_t^2-2}{\epsilon} - 4 - \frac{\pi^2}{4} + 4\,\llog m_t^2-3\,\llog^2 m_t^2 \bigg]\,,\nn\\
    \Gamma^{(2)}_{hh}(p^2=0)\big|^\text{gen.}_{\mathcal{O}(\alpha_s\alpha_t)}&=\frac{\alpha_s m_t^4}{\pi^3v^2}\bigg[-\frac{9}{2\epsilon^2}+\frac{9\,\llog m_t^2}{\epsilon}-4-\frac{3\pi^2}{4}-9\,\llog^2m_t^2\bigg]\,,\nn\\
    \Gamma^{(2)}_{hhh}(0,0,0)\big|^\text{gen.}_{\mathcal{O}(\alpha_s\alpha_t)}&=\frac{\alpha_s m_t^4}{\pi^3v^3}\bigg[-\frac{9}{\epsilon^2} + \frac{18}{\epsilon} (1 + \llog m_t^2)-8 - \frac{3 \pi^2}{2} - 18 \,\llog m_t^2 (2 + \llog m_t^2)\bigg]\,,\nn\\
    \Gamma^{(2)}_{hhhh}(0,0,0,0)\big|^\text{gen.}_{\mathcal{O}(\alpha_s\alpha_t)}&=\frac{\alpha_s m_t^4}{\pi^3v^4}\bigg[-\frac{9}{\epsilon^2} + \frac{18}{\epsilon} (3 + \llog m_t^2)-80 - \frac{3 \pi^2}{2} - 108\, \llog m_t^2 - 18\, \llog^2m_t^2\bigg]\,,
\end{align}
while at $\mathcal{O}(\alpha_t^2)$ we have
\begin{align}
    \Gamma^{(2)}_h\big|^\text{gen.}_{\mathcal{O}(\alpha_t^2)}&=\ \frac{3m_t^6}{16\pi^4v^3}\bigg[\frac{3}{4 \epsilon^2} + \frac{1}{\epsilon}\bigg(2 - \frac{3}{2}\,\llog m_t^2\bigg)+\frac{9}{2} + \frac{5 \pi^2}{24} - 4 \,\llog m_t^2 + \frac{3}{2}\,\llog^2 m_t^2\bigg]\,,\nn\\
    \Gamma^{(2)}_{hh}(p^2=0)\big|^\text{gen.}_{\mathcal{O}(\alpha_t^2)}&=\ \frac{3m_t^6}{16\pi^4v^4}\bigg[\frac{9}{4 \epsilon^2} + \frac{3}{\epsilon}\bigg(1 - \frac{3}{2}\,\llog m_t^2\bigg)+\frac{11}{2} + \frac{5 \pi^2}{8} - 6 \,\llog m_t^2 + \frac{9}{2} \,\llog^2 m_t^2\bigg]\,,\nn\\
    \Gamma^{(2)}_{hhh}(0,0,0)\big|^\text{gen.}_{\mathcal{O}(\alpha_t^2)}&=\ \frac{3m_t^6}{16\pi^4v^5}\bigg[\frac{9}{2 \epsilon^2} - \frac{3}{\epsilon}\bigg(1 + 3\,\llog m_t^2\bigg)-1 + \frac{5 \pi^2}{4} + 6\, \llog m_t^2 + 9 \,\llog^2 m_t^2\bigg]\,,\nn\\
    \Gamma^{(2)}_{hhhh}(0,0,0,0)\big|^\text{gen.}_{\mathcal{O}(\alpha_t^2)}&=\ \frac{3m_t^6}{16\pi^4v^6}\bigg[\frac{9}{2 \epsilon^2} - \frac{3}{\epsilon}\bigg(7 + 3\,\llog m_t^2\bigg)+11 + \frac{5 \pi^2}{4} + 42\, \llog m_t^2 + 9\, \llog^2m_t^2\bigg]\,.
\end{align}
In the above equations, we use the shorthand notation $\llog x\equiv \ln x/Q^2$, where $Q$ is the renormalisation scale.

To these expressions, we must add contributions from subloop renormalisation, which in general form read
{\allowdisplaybreaks
\begin{align*}
    \Gamma^{(2)}_h\big|^\text{subloop}&=\ \frac{3m_t^2}{8 \pi^2 v}\bigg[\frac{\olct v^2}{v^2} - \olct Z_h \bigg]\mathbf{A}_0(m_t^2)\nn\\
    &\quad-\frac{3 m_t \olct m_t}{2 \pi^2 v} \Big[\mathbf{A}_0(m_t^2) + m_t^2 \mathbf{B}_0(0, m_t^2, m_t^2)\Big]\,,\nn\\
    \Gamma^{(2)}_{hh}(p^2=0)\big|^\text{subloop}&=\ \frac{3m_t^2}{4 \pi^2 v^2}\bigg[\frac{\olct v^2}{v^2} - \olct Z_h \bigg]\Big[\mathbf{A}_0(m_t^2) + 2 m_t^2 \mathbf{B}_0(0, m_t^2, m_t^2)\Big]\nn\\
    &\quad-\frac{3 m_t \olct m_t}{2 \pi^2 v^2} \bigg[\mathbf{A}_0(m_t^2) + m_t^2 \Big(5 \mathbf{B}_0(0, m_t^2, m_t^2) + 4 m_t^2 \mathbf{C}_0(m_t^2, m_t^2, m_t^2)\Big)\bigg] \,,\nn\\
    \Gamma^{(2)}_{hhh}(0,0,0)\big|^\text{subloop}&=\ \frac{9m_t^4}{4 \pi^2 v^3}\bigg[\frac{\olct v^2}{v^2} - \olct Z_h \bigg] 
    \Big[3\mathbf{B}_0(0, m_t^2, m_t^2) + 4 m_t^2 \mathbf{C}_0(m_t^2, m_t^2,m_t^2)\Big]\nn\\
    &\quad-\frac{18m_t^3\olct m_t}{\pi^2 v^3}\bigg[\mathbf{B}_0(0, m_t^2, m_t^2) + m_t^2 \Big(3\mathbf{C}_0(m_t^2, m_t^2,m_t^2)\nn\\
    &\hspace{4cm}+ 2m_t^2 \mathbf{D}_0(m_t^2, m_t^2, m_t^2, m_t^2)\Big)\bigg]\,,\nn\\
    \Gamma^{(2)}_{hhhh}(0,0,0,0)\big|^\text{subloop}&=\ \frac{9m_t^4}{\pi^2 v^4}\bigg[\frac{\olct v^2}{v^2} - \olct Z_h \bigg] \nn\\
    &\qquad\times\bigg[\mathbf{B}_0(0, m_t^2, m_t^2) + 8 m_t^2 \Big(\mathbf{C}_0(m_t^2, m_t^2,m_t^2) + m_t^2 \mathbf{D}_0(m_t^2, m_t^2, m_t^2, m_t^2)\Big)\bigg]\nn\\
    &\quad-\frac{18 m_t^3 \olct m_t}{\pi^2 v^4}\times\bigg\{\mathbf{B}_0(0, m_t^2, m_t^2) + m_t^2 \bigg[13 \mathbf{C}_0(m_t^2, m_t^2, m_t^2)\nn\\
    &\hspace{4cm} + 4 m_t^2\Big(7 \mathbf{D}_0(m_t^2,m_t^2,m_t^2,m_t^2)\nn\\
    &\hspace{4cm}+ 4m_t^2 \mathbf{E}_0(m_t^2,m_t^2,m_t^2,m_t^2, m_t^2)\Big)\bigg]\bigg\}\,,
\end{align*}
}where $\mathbf{A}_0$, $\mathbf{B}_0$, $\mathbf{C}_0$, $\mathbf{D}_0$, $\mathbf{E}_0$ denote standard, unrenormalised, one-loop Passarino-Veltmann functions~\cite{Passarino:1978jh}, and where we have omitted momenta arguments for $\mathbf{C}_0$, $\mathbf{D}_0$, and $\mathbf{E}_0$ as these are all set to zero. We obtain the different counterterms by using the functionalities of \texttt{FormCalc}~\cite{Hahn:1998yk,Hahn:2016ebn} and the default SM model file for \texttt{FeynArts}. Moreover, for the present calculation, we only need contributions of $\mathcal{O}(\alpha_s)$ (for $\olct m_t$ only) and of $\mathcal{O}(\alpha_t)$.

The particular cases of the Passarino-Veltmann functions in the above equation can be simplified using the relations
\begin{align}
    \mathbf{C}_0(x,x,x)=\ \frac{1}{2}\frac{\partial}{\partial x}\mathbf{B}_0(0,x,x)\,,\quad
    &\mathbf{D}_0(x,x,x,x)=\ \frac{1}{6}\frac{\partial^2}{\partial x^2}\mathbf{B}_0(0,x,x)\,,\nn\\
    \mathbf{E}_0(x,x,x,x,x)=&\ \frac{1}{24}\frac{\partial^3}{\partial x^3}\mathbf{B}_0(0,x,x)\,.
\end{align}

\paragraph*{\underline{\MS renormalisation}:}
Starting with an \MS renormalisation scheme, we have
\begin{align}
    \frac{1}{m_t}\olct m_t^\MS&=\frac{1}{\epsilon}\bigg[-\frac{\alpha_s}{\pi} + \frac{3 m_t^2}{32 \pi^2 v^2}\bigg]\,,\nn\\
    \olct Z_h^\MS=\left(\frac{\olct v^2}{v^2}\right)^\MS&=-\frac{3m_t^2}{8\pi^2v^2\epsilon}\,.
\end{align}
Additionally, genuine two-loop counterterms for $\lambda_{hhh}$ and $\lambda_{hhhh}$ are also required. However, in the SM (as well as in aligned BSM models) it can be shown that such a pure two-loop counterterm is fixed entirely\footnote{We note that terms involving the two-loop counterterm for the EW vacuum expectation value $v$ do not contribute at $\mathcal{O}(\alpha_s\alpha_t)$ or $\mathcal{O}(\alpha_t^2)$, and therefore do not appear here.} in terms of the two-loop tadpole and mass counterterms --- adapting for instance the discussion in Ref.~\cite{Kanemura:2004mg} to the two-loop order. It follows that the combinations 
\begin{align}
   &\Gamma^{(2)}_{hhh}(0,0,0)\big|^\text{gen.}+\Gamma^{(2)}_{hhh}(0,0,0)\big|^\text{subloop}\nn\\
   &-\frac{3}{v}\bigg[\Gamma^{(2)}_{hh}(0)\big|^\text{gen.}+\Gamma^{(2)}_{hh}(0)\big|^\text{subloop}-\frac{1}{v}\bigg(\Gamma^{(2)}_{h}\big|^\text{gen.}+\Gamma^{(2)}_{h}\big|^\text{subloop}\bigg)\bigg]
\end{align}
and
\begin{align}
   &\Gamma^{(2)}_{hhhh}(0,0,0,0)\big|^\text{gen.}+\Gamma^{(2)}_{hhhh}(0,0,0,0)\big|^\text{subloop}\nn\\
   &-\frac{3}{v^2}\bigg[\Gamma^{(2)}_{hh}(0)\big|^\text{gen.}+\Gamma^{(2)}_{hh}(0)\big|^\text{subloop}-\frac{1}{v}\bigg(\Gamma^{(2)}_{h}\big|^\text{gen.}+\Gamma^{(2)}_{h}\big|^\text{subloop}\bigg)\bigg]
\end{align}
are UV-finite, and correspond to the Higgs mass and tadpole being renormalised \emph{on-shell}. These two expressions --- including the finite parts of the one- and two-point functions --- can then be identified, up to a minus sign, with the two-loop corrections to the trilinear and quartic Higgs couplings (\ie $-\delta^{(2)}\lambda_{hhh}$ and $-\delta^{(2)}\lambda_{hhhh}$), respectively. 

Combining all expressions above, we then find 
\begin{align}
    \left.\delta^{(2)}\lambda_{hhh}\right|^\MS=&\ \frac{2\alpha_s m_t^4}{\pi^3v^3} \big[1 + 6 \llog m_t^2\big] - \frac{3 m_t^6 }{16 \pi^4 v^5}\big[6 \llog m_t^2-7\big]\,,\nn\\ 
    \left.\delta^{(2)}\lambda_{hhhh}\right|^\MS=&\  \frac{16\alpha_s m_t^4}{\pi^3v^4} \big[2 + 3 \llog m_t^2\big]-\frac{3 m_t^6 }{2 \pi^4 v^6}\big[3 \llog m_t^2-2\big] \,,
\end{align}
in terms of \MS-renormalised parameters $m_t$ and $v$, and with the Higgs wave-function renormalisation (WFR) \MS as well. These results correspond exactly to Eqs.\ (V.2) and (V.3) in \ccite{Braathen:2019zoh}. 

\paragraph*{\underline{On-shell renormalisation}:} Next, one can also consider a full on-shell (OS) renormalisation scheme. The OS-scheme counterterms of the top-quark mass and the Higgs WFR can be found to be, up to $\mathcal{O}(\epsilon^0)$, 
\begin{align}
    \frac{1}{m_t}\olct m_t^\text{OS}=&\ \frac{\alpha_s}{\pi}\bigg[-\frac{1}{\epsilon} + \llog m_t^2-\frac{4}{3}\bigg]
    +\frac{3m_t^2}{32\pi^2v^2}\bigg[\frac{1}{\epsilon} +\frac{8}{3} - \llog m_t^2 \bigg]\,,\nn\\
    \olct Z_h^\text{OS}=&\ \frac{3m_t^2}{8\pi^2v^2}\bigg[-\frac{1}{\epsilon}+\frac{2}{3} + \llog m_t^2 \bigg]\,.
\end{align}
Depending on the choice of input scheme for electroweak parameters, the finite part of the corresponding counterterm of the EW vacuum expectation value (VEV) takes different forms. A first choice (following Ref.~\cite{Braathen:2019pxr,Braathen:2019zoh}) is to relate $v$ to the Fermi constant $G_F$, which results in
\begin{align}
    \left(\frac{\olct v^2}{v^2}\right)^{\text{OS, }G_F}&=\frac{3m_t^2}{16\pi^2v^2} \bigg[-\frac{2}{\epsilon}+2 \llog m_t^2 - 1\bigg]\,.
\end{align}
Another possibility is instead to relate the EW VEV to $M_W$, $M_Z$, and the electric charge $e$, and in this case, the OS counterterm for $v$ reads
\begin{align}
    \left(\frac{\olct v^2}{v^2}\right)^{\text{OS, }\alpha_\text{em}}&= \frac{\olct (M_W^2)^\text{OS}}{M_W^2}+\frac{c_W^2}{s_W^2}\bigg(\frac{\olct (M_Z^2)^\text{OS}}{M_Z^2}-\frac{\olct (M_W^2)^\text{OS}}{M_W^2}\bigg)-\frac{2\olct e^\text{OS}}{e}=\nn\\
    &=\frac{3m_t^2}{16\pi^2v^2} \bigg[-\frac{2}{\epsilon}+2 \llog m_t^2 - 1-\frac{c_W^2}{s_W^2}\bigg]\,.
\end{align}
With the first choice, we find
\begin{align}
   \left.\delta^{(2)}\lambda_{hhh}\right|^{\text{OS},\ G_F}=&\ \frac{18\,\alpha_s M_t^4}{\pi^3v_\text{OS}^3}  - \frac{117 M_t^6}{32 \pi^4 v_\text{OS}^5}\,,\nn\\ 
   \left.\delta^{(2)}\lambda_{hhhh}\right|^{\text{OS},\ G_F}=&\  \frac{96\,\alpha_s M_t^4}{\pi^3v_\text{OS}^4} -\frac{39 M_t^6}{2 \pi^4 v_\text{OS}^6} \,,
\end{align}
in terms of the top-quark pole mass $M_t$ and of $v_{\text{OS}}\equiv (\sqrt{2}G_F)^{-1/2}\simeq 246.22$ GeV. These expressions agree exactly with those in Eq.~(V.6) of \ccite{Braathen:2019zoh}. In the second case, we obtain the following new results
\begin{align}
    \left.\delta^{(2)}\lambda_{hhh}\right|^{\text{OS},\ \alpha_\text{em}}=&\ \frac{18\,\alpha_s M_t^4}{\pi^3\bar{v}_\text{OS}^3}  - \frac{M_t^6}{32 \pi^4 \bar{v}_\text{OS}^5}\left[117+\frac{27\,c_W^2}{s_W^2}\right]\,,\nn\\ 
    \left.\delta^{(2)}\lambda_{hhhh}\right|^{\text{OS},\ \alpha_\text{em}}=&\  \frac{96\,\alpha_s M_t^4}{\pi^3\bar{v}_\text{OS}^4} -\frac{M_t^6}{2 \pi^4 \bar{v}_\text{OS}^6}\left[39+\frac{9\,c_W^2}{s_W^2}\right] \,,
\end{align}
where the same top-quark pole mass $M_t$ is employed but with $\bar{v}_\text{OS}\equiv \frac{2M_Ws_W}{e}\simeq 250.69$ GeV as well as $e\equiv \sqrt{4\pi \alpha_\text{em}(0)}$ and $c_W\equiv M_W/M_Z$ is the cosine of the weak mixing angle.

\begin{figure}
    \centering
\includegraphics[width=0.8\textwidth]{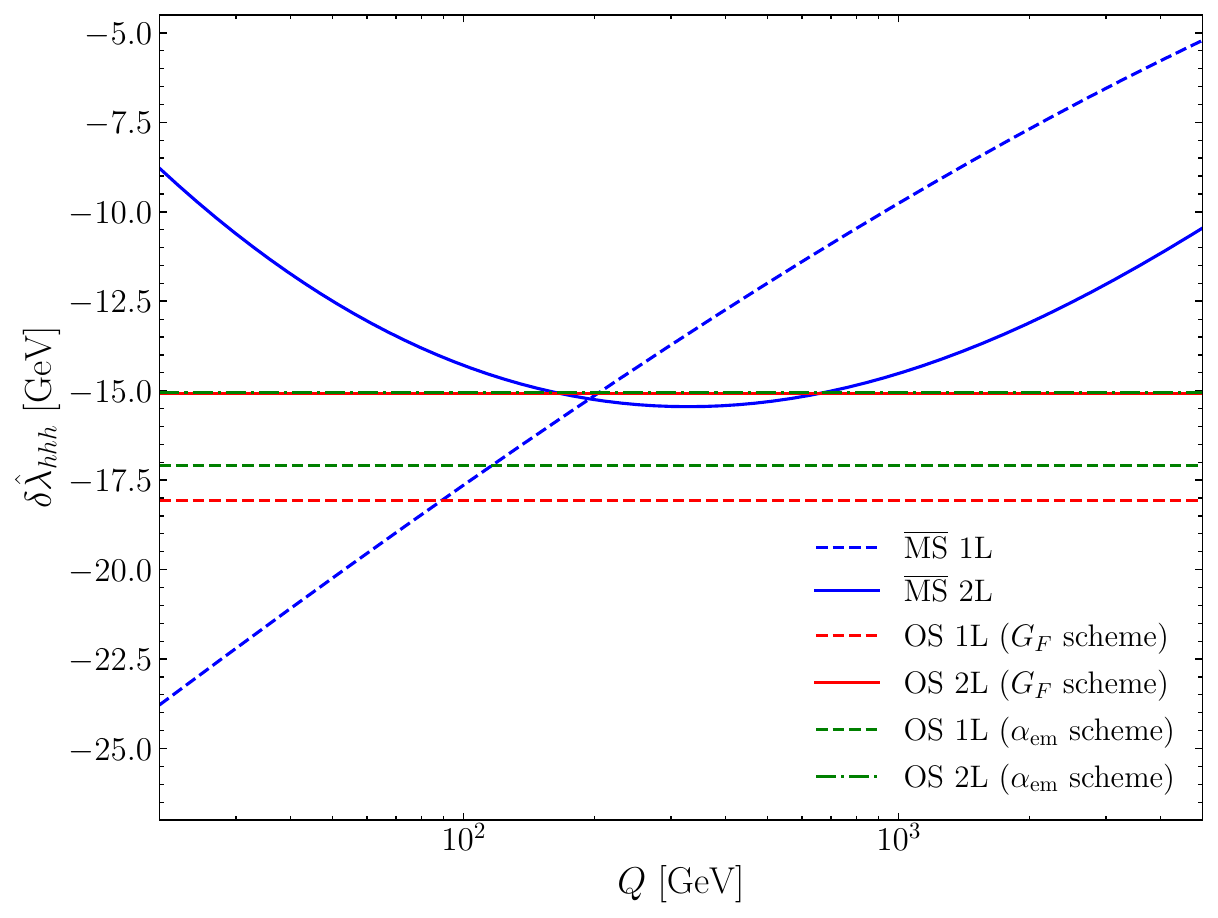}
    \caption{Leading one- and two-loop corrections to the trilinear Higgs coupling in the Standard Model. Blue curves correspond to the \MS scheme, while the red and green ones correspond to two versions of OS scheme, with the EW VEV related respectively to $G_F$ (red) and to $M_W$, $M_Z$ and $\alpha_\text{em}(0)$ (green).}
    \label{fig:hhh_SM_2L}
\end{figure}

These results\footnote{Our numerical inputs are taken to be $G_F=1.1663788\cdot 10^{-5}\text{ GeV}^{-2}$, $M_W=80.3692\text{ GeV}$, $M_Z=91.1880\text{ GeV}$, $\alpha_\text{em}(0)=1/137.035999084$, $\alpha_s(M_Z)=0.118$, $M_t=172.57\text{ GeV}$ --- see Ref.~\cite{ParticleDataGroup:2024cfk}.} are illustrated in \cref{fig:hhh_SM_2L}, which compares results for the predictions for the loop corrections, denoted $\delta\hat\lambda_{hhh}$, to $\hat\lambda_{hhh}$  at leading one- (dashed lines) and two-loop (solid or dot-dashed lines) orders (we note that, following the choice made in~\ccite{Braathen:2019zoh,Braathen:2019pxr}, we do not include external-leg corrections in the \MS results for $\hat\lambda_{hhh}$). Results in the three renormalisation schemes discussed above --- \MS, OS with the EW VEV related to $G_F$, and OS with the EW VEV related to $M_W$, $M_Z$ and $\alpha_\text{em}(0)$ --- are indicated by the blue, red, and green colours, respectively. As expected, both versions of OS schemes are manifestly independent of the renormalisation scale. At one loop, the predictions for $\delta\hat\lambda_{hhh}$ in the two OS schemes differ by about $970$ MeV, while at two loops this difference is reduced to $42$ MeV --- which can be used as an estimate of three-loop mixed QCD--EW corrections, which are not included in this calculation.


\subsection{Cross-checks in the \texorpdfstring{$\mathbb{Z}_2$-symmetric}{Z2-symmetric} singlet extension of the SM}
\label{sec:SSMZ2}

As a first example of a BSM model, we investigate in this Section the application of our generic results in the context of the $\Ztwo$-symmetric singlet extension of the SM (\zssm). 

\subsubsection{The model}

The scalar sector of the \zssm~contains the usual SM-like doublet $\Phi$, decomposed as 
\begin{equation}
    \Phi = \frac{1}{\sqrt{2}} \begin{pmatrix} \sqrt{2}G^+ \\ v+h+i G \end{pmatrix}\,,
\end{equation}
and a singlet $S$ that is charged under an unbroken global $\mathbb{Z}_2$ symmetry. In terms of these two states, the tree-level scalar potential of the \zssm~can be written as 
\begin{equation}
    V_{\Ztwo}(\Phi,S) = 
    \mu^2|\Phi|^2 + 
    \frac{\lambda_H}{2} |\Phi|^4 + 
    \frac{\mu_S^2}{2} S^2 + 
    \frac{\lambda_S}{2} S^4 +
    \frac{\lambda_{SH}}{2}S^2|\Phi|^2\,.
    \label{eq:SSMZ2:pot}
\end{equation}
The minimisation condition of the tree-level potential reads
\begin{align}
    0=\frac{1}{v}\frac{\partial V_{\Ztwo}}{\partial h}\bigg|_{h=0}=\frac{t_h^{(0)}}{v}=\mu^2+\frac12\lambda_Hv^2\,,
\end{align}
while the tree-level mass eigenstates are 
\begin{align}
 m_h^2=\mu^2+\frac32\lambda_Hv^2\,,\quad\text{and}\quad m_S^2 =\mu_S^2+\frac12\lambda_{SH}v^2\,.   
\end{align}
These relations allow us to rewrite, at the tree level, the quartic couplings $\lambda_H$ and $\lambda_{SH}$ as
\begin{align}
\label{eq:Z2SSM_repquartics}
    \lambda_H = \frac{m_h^2}{v^2}-\frac{t_h^{(0)}}{v^3}\,,\quad\text{and} \quad \lambda_{SH} = \frac{2(m_S^2-\mu_S^2)}{v^2}\,.
\end{align}
We consider the limit $m_S\gg m_h$ such that $m_h\approx 0$ and consequently $\lambda_H\approx 0$ are good approximations beyond tree-level. We turn now to the discussion of the counterterms required, in this approximation, to obtain a UV-finite result for $\lambda_{hhh}$ at two loops, expressed entirely in terms of OS quantities.
First, for the subloop renormalisation we require the following conditions and counterterms: 
\begin{itemize}
    \item an OS renormalisation of the tree-level masses and the scalar fields:
        \begin{equation}
            \delta^{(1)}_\text{CT} m_{h}^2 = \Sigma^{(1)}_{hh}(p^2=0)\,,\quad \text{and} \quad \delta^{(1)}_\text{CT} m_{S}^2 = \Sigma^{(1)}_{SS}(p^2=m_{S}^2)\,,
        \end{equation}
     and
    \begin{equation}
       \delta^{(1)}_\text{CT} Z_{h} = - \left. \frac{\partial}{\partial p^2}\Sigma^{(1)}_{hh}(p^2)\right|_{p^2=0}\,, \quad \text{and} \quad \delta^{(1)}_\text{CT} Z_{S} = - \left. \frac{\partial}{\partial p^2}\Sigma^{(1)}_{SS}(p^2)\right|_{p^2=m_S^2}\,,
    \end{equation}
    
    \item an OS renormalisation of the tadpole $\hat{t}_h$:
    \begin{equation}
        \hat{t}_h = t_h^{(0)} + t_h^{(1)} + \delta^{(1)}_\text{CT} t_h = 0 \Rightarrow \delta^{(1)}_\text{CT} t_h = - t_h^{(1)},\,
    \end{equation}
    and finally,
    \item an ``OS-like'' (or ``decoupling inspired'') renormalisation of $\mu_S$, demanding the correct decoupling behaviour of corrections involving the singlet self-coupling, following \ccite{Braathen:2019zoh}:
    \begin{equation}
        \delta^{(1)}_\text{CT} \mu_S^2 = \frac{3 \lambda_S \mu_S^2}{8\pi^2m_S^2} \textbf{A}_0(m_S^2)
    \end{equation}
    where $\textbf{A}_0$ is the (unrenormalised) scalar one-loop one-point function~\cite{Passarino:1978jh}.
\end{itemize}
To cancel all divergences at the two-loop level and to ensure a proper OS description, we employ the following two-loop counterterms:
\begin{itemize}
    \item mass counterterm of the SM-like Higgs boson:
        \begin{equation}
            \delta^{(2)}_\text{CT} m_{h}^2 = \Sigma^{(2)}_{hh}(p^2=0) + \frac{1}{2}\delta^{(1)}_\text{CT}Z_{hh} \delta^{(1)}_\text{CT}m_h^2 \,, 
        \end{equation}
    \item two-loop tadpole:
    \begin{equation}
        \delta^{(2)}_\text{CT} t_h = - t_h^{(2)} -  \delta^{(1)}_\text{CT}Z_{hh} \delta^{(1)}_\text{CT}t_h \,.
    \end{equation}
\end{itemize}
With these ingredients, we can construct the renormalised trilinear coupling at one loop
\begin{equation}
   \delta^{(1)}\lambda_{hhh} =  \delta^{(1)}_\text{diag.}\lambda_{hhh} + \frac{3}{v^2}\left(\delta^{(1)}_\text{CT}t_h - v\,\delta^{(1)}_\text{CT}m_h^2 \right)\,,
\end{equation}
and at two loops
\begin{equation}
    \delta^{(2)}\lambda_{hhh} = \frac{3}{2}\delta^{(1)}\lambda_{hhh}\left(\delta^{(1)}_\text{CT}Z_{hh} + \Sigma_{hh}^{\prime}(0)\right)  + \delta^{(2)}_\text{diag.}\lambda_{hhh}+ \frac{3}{v^2}\left(\delta^{(2)}_\text{CT}t_h - v\,\delta^{(2)}_\text{CT}m_h^2 \right)\,,
    \label{eq:SSMZ2OS2L}
\end{equation}
where the subscript ``diag'' denotes the sum of all Feynman-diagrammatic contributions, including one-loop diagrams with one-loop mass- and vertex-counterterm insertions (that stem from the subloop renormalisation). In \cref{eq:SSMZ2OS2L} we introduce the notation $\Sigma^{\prime}(x)=\frac{\partial}{\partial p^2}\Sigma(p^2)\big|_{p^2=x}$.

\subsubsection{Leading two-loop corrections to the trilinear Higgs coupling}

The renormalisation procedure described in the previous paragraph combined with the diagrammatic calculation of all one- and two-loop diagrams contributing to one-, two- and three-point functions of the Higgs boson using the generic results from \cref{sec:generic_calc,sec:FA_mapping} leads to exact agreement with the results obtained for the leading BSM contributions to $\lambda_{hhh}$ in the \zssm~in \ccite{Braathen:2019zoh}. In contrast to our results, in \ccite{Braathen:2019zoh} the corrections to the trilinear coupling were obtained by taking field-derivatives of the one- and two-loop effective potential. Furthermore, in this reference, an \MS~renormalisation was performed already at the level of the effective potential; finite counterterm shifts were afterwards included at the level of the derivatives of the effective potential in order to obtain a full OS result for $\lambda_{hhh}$. Therefore, the approach presented in the present work is a strong independent cross-check of both the results in \ccite{Braathen:2019zoh} as well as of the generic results obtained in this work.

First, we obtain for the genuine two-loop corrections to the one-, two-, and three-point functions the following expressions
{\allowdisplaybreaks
\begin{align}
    \Gamma^{(2)}_h\big|^\text{gen.}_{\mathcal{O}(\lambda_S^0)}&=\frac{(m_S^2 - \mu_S^2)^2}{128\pi^4v^3}\bigg[\frac{\mu_S^2-2 m_S^2}{\epsilon^2}+\frac{1}{\epsilon}\big(\mu_S^2-4 m_S^2 + (4 m_S^2 - 2 \mu_S^2) \llog m_S^2\big)\nn\\
    &\qquad\qquad\qquad\quad+\mu_S^2 \left(1 + \frac{\pi^2}{6}\right) - 2 m_S^2 \left(4 + \frac{\pi^2}{6}\right)\nn\\
    &\qquad\qquad\qquad\quad- 2 \llog m_S^2 \big(\mu_S^2-4 m_S^2 + (2 m_S^2 - \mu_S^2)\llog m_S^2\big)\bigg]\,,\nn\\
    \Gamma^{(2)}_h\big|^\text{gen.}_{\mathcal{O}(\lambda_S)}&=\frac{3\lambda_Sm_S^2(m_S^2 - \mu_S^2)}{128\pi^4v}\bigg[-\frac{1}{\epsilon^2} + \frac{
  2 \llog m_S^2-1}{\epsilon} -1 - \frac{\pi^2}{6} - 2 (\llog m_S^2-1) \llog m_S^2\bigg]\,,\nn\\
    \Gamma^{(2)}_{hh}(p^2=0)\big|^\text{gen.}_{\mathcal{O}(\lambda_S^0)}&=\frac{(m_S^2 - \mu_S^2)^2}{128\pi^4v^4}\bigg[\frac{5\mu_S^2-6 m_S^2}{\epsilon^2}-\frac{1}{\epsilon}\bigg(4 m_S^2 + 3 \mu_S^2 - \frac{4 \mu_S^4}{m_S^2} - 2 (6 m_S^2 - 5 \mu_S^2) \llog m_S^2\bigg)\nn\\
    &\hspace{2.5cm}+\frac{4 \mu_S^4}{m_S^2} - 2 m_S^2 \left(4 + \frac{\pi^2}{2}\right) - \mu_S^2 \left(3 - \frac{5\pi^2}{6}\right) \nn\\
    &\hspace{2.5cm}+ 2 \llog m_S^2 \bigg(4 m_S^2 + 3 \mu_S^2 - \frac{4 \mu_S^4}{m_S^2} + (5\mu_S^2-6 m_S^2) \llog m_S^2\bigg)\bigg]\,,\nn\\
    \Gamma^{(2)}_{hh}(p^2=0)\big|^\text{gen.}_{\mathcal{O}(\lambda_S)}&=\frac{3\lambda_S(m_S^2 - \mu_S^2)}{128\pi^4v^2}\bigg[\frac{2\mu_S^2-3 m_S^2}{\epsilon^2} + \frac{m_S^2 - 2 \mu_S^2 + (6 m_S^2 - 4 \mu_S^2) \llog m_S^2}{\epsilon}\nn\\
    &\hspace{3.1cm}+m_S^2 \left(1 - \frac{\pi^2}{2}\right) - 2 \mu_S^2 \left(1 - \frac{\pi^2}{6}\right) \nn\\
    &\hspace{3.1cm}- 2 \llog m_S^2 \Big(m_S^2 - 2 \mu_S^2 + (3 m_S^2 - 2 \mu_S^2) \llog m_S^2\Big)\bigg]\,,\nn\\
    \Gamma^{(2)}_{hhh}(0,0,0)\big|^\text{gen.}_{\mathcal{O}(\lambda_S^0)}&=\frac{4(m_S^2 - \mu_S^2)^3}{128 \pi^4 v^5}\bigg[-\frac{3}{\epsilon^2} + \frac{1}{\epsilon}\bigg(4 - \frac{5 \mu_S^2}{m_S^2} - \frac{2 \mu_S^4}{m_S^4} + 6 \llog m_S^2\bigg)\nn\\
    &\hspace{2.9cm}+\frac{3 \mu_S^2}{m_S^2} - \frac{6 \mu_S^4}{m_S^4} -  \frac{\pi^2}{2} + \bigg(\frac{10 \mu_S^2}{m_S^2} + \frac{4 \mu_S^4}{m_S^4}-8\bigg) \llog m_S^2 - 6 \llog^2 m_S^2\bigg]\,,\nn\\
    \Gamma^{(2)}_{hhh}(0,0,0)\big|^\text{gen.}_{\mathcal{O}(\lambda_S)}&=\frac{3\lambda_S(m_S^2 - \mu_S^2)^2}{128\pi^4v^3}\bigg[-\frac{6}{\epsilon^2} + \frac{1}{\epsilon}\bigg(14 - \frac{8 \mu_S^2}{m_S^2} + 12 \llog m_S^2\bigg)\nn\\
    &\hspace{3.25cm}-2 + \frac{8 \mu_S^2}{m_S^2} - \pi^2 + 4 \bigg(\frac{4 \mu_S^2}{m_S^2} -7\bigg)\llog m_S^2 - 12 \llog^2m_S^2\bigg]\,,
\end{align}
where we have separated terms involving $\lambda_S$ or not, in order to make the expressions more easily readable. As a further example of the impact of our algorithm to reduce the number of diagrams to calculate, we note that in the \zssm, the number of diagrams contributing to the leading BSM corrections to the Higgs three-point function decreases from 57 (at the \texttt{Particles} level) to 18 after reduction.

Next, we require the subloop renormalisation contributions, which are found to be 
\begin{align}
    \Gamma^{(2)}_{h}\big|^\text{subloop}&=\frac{m_S^2-\mu_S^2}{32\pi^2v}\bigg[\bigg( \frac{\olct v^2}{v^2}  + \frac{2 \olct \lambda_{SH}}{\lambda_{SH}}  + \olct Z_h - 2 \olct Z_S \bigg) \mathbf{A}_0(m_S^2) \nn\\
    &\hspace{2.5cm}+ 2 \delta^{(1)}_\text{CT}m_S^2\,  \mathbf{B}_0(0, m_S^2, m_S^2)\bigg]\,,\nn\\
    \Gamma^{(2)}_{hh}(p^2=0)\big|^\text{subloop}&=\frac{m_S^2-\mu_S^2}{16\pi^2v^2}\Bigg\{ \bigg(\frac{\olct\lambda_{SH}}{\lambda_{SH}} + \olct Z_h - \olct Z_S\bigg) \mathbf{A}_0(m_S^2) \nn\\
    &\hspace{-1.5cm}+ \bigg[\olct m_S^2 + 2(m_S^2-\mu_S^2) \bigg(\frac{\olct v^2}{v^2} + \frac{2 \olct\lambda_{SH}}{\lambda_{SH}} + \olct Z_h - 2 \olct Z_S\bigg)\bigg] \mathbf{B}_0(0,m_S^2, m_S^2) \nn\\
    &\hspace{3cm}+ 4 \olct m_S^2 (m_S^2-\mu_S^2)\,\mathbf{C}_0(m_S^2, m_S^2, m_S^2)\Bigg\}\,,\nn\\
    \Gamma^{(2)}_{hhh}(0,0,0)\big|^\text{subloop}&=\frac{3(m_S^2-\mu_S^2)^2}{16\pi^2v^3}\Bigg\{\bigg(4 \frac{\olct\lambda_{SH}}{\lambda_{SH}}+ \frac{\olct v^2}{v^2}+ 3 \olct Z_h - 4 \olct Z_S\bigg) \mathbf{B}_0(0, m_S^2, m_S^2)\nn\\
    &\hspace{-2.5cm}+4 \bigg[ \olct m_S^2 + (m_S^2-\mu_S^2)\bigg( \frac{2\olct \lambda_{SH}}{\lambda_{SH}} + \frac{\olct v^2}{v^2} + \olct Z_h - 2 \olct Z_S\bigg) \bigg] \mathbf{C}_0(m_S^2, m_S^2, m_S^2)\nn\\
    &\hspace{.5cm}+8(m_S^2-\mu_S^2) \olct m_S^2\, \mathbf{D}_0(m_S^2, m_S^2, m_S^2, m_S^2)\Bigg\}\,,
\end{align}
}having here again omitted vanishing momenta in the arguments of $\mathbf{C}_0$ and $\mathbf{D}_0$, as in \cref{sec:SM}. We have expressed these contributions in terms of the counterterm for the portal coupling $\lambda_{SH}$. This can be related to the counterterms presented in the previous Section by differentiating the expression in \cref{eq:Z2SSM_repquartics}, \ie
\begin{align}
    \olct \lambda_{SH} = \frac{2}{v^2}\left(\olct m_S^2-\olct \mu_S^2\right)-\frac{2(m_S^2-\mu_S^2)}{v^2}\frac{\olct v^2}{v^2}\,.
\end{align}

Employing then the counterterm definitions discussed above (and keeping in mind the minus sign between the effective trilinear coupling and the three-point functions), we obtain results for $\delta^{(2)}\lambda_{hhh}$, first in terms of \MS-renormalised BSM parameters $m_S$ and $\mu_S$ (but with OS-renormalised Higgs mass and tadpole)
\begin{align}
\label{eq:Z2SSM_hhh_MSbar}
    \delta^{(2)}\lambda_{hhh}\big|^\MS_{\mathcal{O}(\lambda_S^0)} =&\ \frac{m_S^4}{16\pi^4 v^5} \left(1 - \frac{\mu_S^2}{m_S^2}\right)^4 \big[-m_S^2 - 2 \mu_S^2 + (2 m_S^2 + \mu_S) \llog m_S^2\big]\,,\nn\\
    \delta^{(2)}\lambda_{hhh}\big|^\MS_{\mathcal{O}(\lambda_S)} =&\ \frac{3 \lambda_Sm_S^4 }{32\pi^4 v^3} \left(1 - \frac{\mu_S^2}{m_S^2}\right)^3 \big[1 + 2 \llog m_S^2\big]\,,
\end{align}
where we again split the results between terms involving $\lambda_S$ or not to facilitate reading. Using OS counterterms, we find
\begin{align}
\label{eq:Z2SSM_hhh_OS}
    \delta^{(2)}\lambda_{hhh}\big|^\text{OS}_{\mathcal{O}(\lambda_S^0)} =&\ \frac{3M_S^6}{16\pi^4v^5}\left(1 - \frac{\tilde\mu_S^2}{M_S^2}\right)^4  - \frac{M_S^6}{128\pi^4v^5}\left(1 - \frac{\tilde\mu_S^2}{M_S^2}\right)^5+\frac{21 M_S^4\, M_t^2}{128 \pi^4 v^5}\left(1 - \frac{\tilde\mu_S^2}{M_S^2}\right)^3\nn\\
    &+\frac{3 M_S^2\, M_t^4}{32 \pi^4 v^5}\left(1 - \frac{\tilde\mu_S^2}{M_S^2}\right)^2\,,\nn\\
    \delta^{(2)}\lambda_{hhh}\big|^\text{OS}_{\mathcal{O}(\lambda_S)} =&\ \frac{9\lambda_SM_S^4}{32\pi^4v^3}  \left(1 - \frac{\tilde\mu_S^2}{M_S^2}\right)^3\,,
\end{align}
where $M_S$ and $\tilde\mu_S$ denote the OS-renormalised BSM mass parameters, and $M_t$ is the (OS) top-quark mass. We recall that because $\lambda_S$ only enters the trilinear Higgs coupling at the two-loop order, no counterterm is needed for it. As stated at the beginning of the Section, our \cref{eq:Z2SSM_hhh_MSbar,eq:Z2SSM_hhh_OS} reproduce exactly the results in \ccite{Braathen:2019zoh} (see Eqs.~(V.25) and~(V.31) therein). 

\subsubsection{Matching the \texorpdfstring{$\mathbb{Z}_2$SSM}{Z2SSM} onto the SM}
Another result from the literature that we can reproduce in the $\mathbb{Z}_2$SSM is the dominant two-loop threshold corrections to the SM-like quartic Higgs coupling $\lambda_H$~\cite{Braathen:2018htl} when integrating out the singlet $S$ and matching the $\mathbb{Z}_2$SSM onto the SM. This illustrates the fact that our results can also be applied to the derivation of two-loop matching conditions for renormalisable scalar operators, which are a crucial ingredient for \eg precision Higgs boson mass predictions~\cite{Slavich:2020zjv}. For this calculation, 348 Feynman diagrams are generated by \FeynArts, which are subsequently reduced to 38 with the algorithm described in \cref{sec:diagram_reduction}.

Keeping our expressions here in terms of $\lambda_{SH}$ (\ie we do not replace this coupling using \cref{eq:Z2SSM_repquartics}), the leading genuine two-loop contributions --- of $\mathcal{O}(\lambda_{SH}^3)$ and $\mathcal{O}(\lambda_{SH}^2\lambda_S)$ --- to the four-point function $\Gamma_{hhhh}$ read (omitting all momenta arguments, as they are set to zero)
\begin{align}
    (4\pi)^4\Gamma^{(2)}_{hhhh}\big|^\text{gen.}_\text{\zssm}=&-\frac{3 \lambda _{SH}^2(\lambda_{SH}+3\lambda_S)}{ \epsilon^2}+\frac{3 \lambda_{SH}^2\big[\lambda_{SH}(2 \llog m_S^2-1)+3\lambda_S \left(2 \llog m_S^2+1\right)\big]}{\epsilon}\nn\\
    &-\lambda_{SH}^3\left(6 \llog^2 m_S^2-6 \llog  m_S^2+\frac{\pi^2}{2}+3\right)\nn\\
    &-3 \lambda_{SH}^2 \lambda_S\left(6 \llog^2 m_S^2+6 \llog  m_S^2+\frac{\pi^2}{2}-3\right)\,.
\end{align}
The corresponding subloop renormalisation contributions are found to be 
\begin{align}
   (4\pi)^2\Gamma^{(2)}_{hhhh}\big|^\text{subloop}_\text{\zssm}=3 \lambda_{SH}^2\bigg[&\left(\frac{\delta^{(1)}_\text{CT}\lambda_{SH}}{\lambda_{SH}}+\delta^{(1)}Z_h-\delta^{(1)}Z_S\right) \mathbf{B}_0(0, m_S^2, m_S^2) \nn\\
   &+ \delta^{(1)}_\text{CT}m_S^2 \mathbf{C}_0(m_S^2,m_S^2,m_S^2)\bigg]\,,
\end{align}
while the genuine two-loop counterterm for $\lambda_H$ does not contribute at the order we are considering. 

For the purpose of this matching calculation, we employ an \MS renormalisation scheme for $\lambda_{SH}$ and $m_S^2$, and we work in the unbroken phase of the theory (\ie in the limit $v\to0$). At leading order in powers of $\lambda_{SH},\lambda_S\gg \lambda_H$, we then find
\begin{align}
    \olct Z_h^\MS=&\ \delta^{(1)}_\text{CT}Z_S^\MS=0\,,\nn\\
    \olct\lambda_{SH}^\MS=&\ \frac{1}{8\pi^2\epsilon}\lambda_{SH} (\lambda_{SH} + 3 \lambda_S)\,,\nn\\
    \olct (m_S^2)^\MS=&\ \frac{3}{8\pi^2\epsilon} \lambda_S m_S^2 \,.
\end{align}
Combining all these elements, we obtain the leading two-loop $\mathcal{O}(\lambda_{SH}^3)$ and $\mathcal{O}(\lambda_{SH}^2\lambda_S)$ threshold corrections to the SM quartic Higgs coupling\footnote{We use as our convention that the SM quartic coupling $\lambda$ is defined as $\mathcal{L}\supset -\frac{1}{2}\lambda |\Phi|^4$, where $\Phi$ is the Higgs doublet.} $\lambda$ from integrating out the singlet state
\begin{align}
    \Delta^{(2)}\lambda=&-\frac13\big[\Gamma^{(2)}_{hhhh}\big|^\text{gen.}_\text{\zssm}+\Gamma^{(2)}_{hhhh}\big|^\text{subloop, \MS}_\text{\zssm}\big]\nn\\
    =&\ \frac{\lambda_{SH}^2}{256\pi^4} \bigg[\lambda_{SH} \big(\llog m_S^2-1\big)^2 + 3 \lambda_S \big(\llog^2m_S^2+\llog m_S^2-1\big)\bigg]\,,
\end{align}
which corresponds exactly to Eq.~(29) of Ref~\cite{Braathen:2017jvs} once one takes the limits of $\lambda\to0$ and $v\to0$ therein. We note that as there are no SM contributions at the considered orders, there is no need to subtract SM expressions to obtain $\Delta^{(2)}\lambda$.


\subsection{Matching the SM and the THDM: two-loop corrections to the quartic Higgs coupling}

In this Section, we turn to a more complex BSM model and calculate the two-loop threshold correction for the quartic Higgs coupling between the SM and the Two-Higgs-Doublet Model (THDM) taking into account all contributions of \order{\alt^2}. Following \ccite{Bahl:2018jom,Bahl:2020jaq}, we work in a general THDM in which not only the second doublet couples to top quarks --- with the coupling $h_t$ --- but also the first doublet --- with the coupling $h_t^\prime$ (with both $h_t$ and $h_t^\prime$ being assumed to be real).

As in \ccite{Bahl:2020jaq}, we perform our calculation in the unbroken phase --- \ie, we set the VEV to zero. This greatly simplifies the calculation, since no mixing between the heavy Higgs bosons needs to be considered. We do, however, retain a finite top-quark mass in order to regulate infrared singularities, once again like what was done in \ccite{Bahl:2020jaq}. 

Generating all $\mathcal{O}(\alpha_t^2)$ diagrams with \FeynArts, we obtain 480 Feynman diagrams at the classes level which we reduced to only 48 unique two-loop diagrams with appropriate symmetry factors by using the canonical edges formalism. The result (with all external momenta set to zero) is given by
\begin{align}
    (4\pi)^4\Gamma^{(2)}_{hhhh}\big|^\mathrm{gen.}_\text{THDM}  
    =&\ (c_\beta h_t-s_\beta h_t^\prime)^2\,(s_\beta h_t+ c_\beta h_t^\prime)^4 \nn\\
     &\times\bigg[
    \frac{1}{\epsilon^2}\frac{27}{c_\beta^2} - \frac{1}{\epsilon}\frac{126 + 54 \llog m_t^2}{c_\beta^2} - 27 \llog^2 M_{H^+}^2 + 108 \llog M_{H^+}^2 \nn\\
    & \hspace{.4cm}+ 54 \llog M_{H^+} \llog m_t^2 + 18 \frac{11 - 3 c_{2\beta}}{c_\beta^2} \llog m_t^2 \nn\\
    & \hspace{.4cm}- \frac{27}{2 c_\beta^2} (3 - c_{2\beta}^2) \llog^2 m_t^2 + \frac{3}{2c_\beta^2}\pi\left(1 - 4 c_{2\beta} \right) + 6\frac{13 + 2 c_{2\beta}}{c_\beta^2}\bigg]\;,
        \label{eq:SM2THDM_THDMgen}
\end{align}
where $M_{H^+}$ is the mass scale of the heavy Higgs bosons.

The additional contribution from the \MS subloop renormalisation reads
\begin{align}
(4\pi)^4 \Gamma^{(2)}_{hhhh}\big|^\mathrm{subloop}_\text{THDM} 
    =&\ (c_\beta h_t-s_\beta h_t^\prime)^2\,(s_\beta h_t+ c_\beta h_t^\prime)^4\,\frac{1}{c_\beta^2} \nn\\
    & \times \bigg[ - \frac{54}{\epsilon^2} + \frac{9}{\epsilon}\left(19 + 6 \llog m_t^2\right) - 27 \llog^2 m_t^2 - 171\llog m_t^2 \nn \\
    & \hspace{.45cm}- 144 - \frac{9}{2}\pi^2\bigg]\;.
        \label{eq:SM2THDM_THDMsub}
\end{align}
The corresponding genuine and subloop SM corrections of $\mathcal{O}(\alpha_t^2)$ are given by
\begin{align}
    (4\pi)^4 \Gamma_{hhhh}^{(2)}\big|^\mathrm{gen.}_\text{SM}  
    = y_t^6\bigg[
    & \frac{27}{\epsilon^2} - \frac{18}{\epsilon}\left(7 + 3 \llog m_t^2\right) + 54 \llog^2 m_t^2 + 252 \llog m_t^2 + 66\bigg] \label{eq:SM2THDM_SMgen}
\end{align}
and
\begin{align}
    (4\pi)^4 \Gamma_{hhhh}^{(2)}\big|^\mathrm{subloop}_\text{SM} 
    = y_t^6 \bigg[
    & - \frac{54}{\epsilon^2} + \frac{9}{\epsilon}\left(19 + 6 \llog m_t^2\right) - 27 \llog^2 m_t^2 - 171\llog m_t^2 \nn\\
    & - 144 - \frac{9}{2}\pi^2\bigg]\;,
        \label{eq:SM2THDM_SMsub}
\end{align}
respectively. Here, $y_t$ is the SM top-Yukawa coupling.

The relation between the SM and the THDM Yukawa couplings, which receives a one-loop threshold correction, reads~\cite{Bahl:2018jom}
\begin{align}
    y_t(Q) = (s_\beta h_t + c_\beta h_t^\prime)\left[1 - \frac{1}{(4\pi)^2}\frac{3}{8}(c_\beta h_t - s_\beta h_t^\prime)^2\left(1 - 2 \llog M_{H^+}^2\right)\right]\;.
\end{align}
Using this relation to parametrise the one- and two-loop SM corrections in terms of $h_t$ and $h_t^\prime$, and taking the difference between the THDM and SM two-loop corrections, we obtain the two-loop shift to the SM quartic Higgs coupling $\lambda$ at the matching scale
\begin{align}
    \Delta^{(2)}\lambda
    & =
        -\frac{1}{3}\left[ \Gamma_{hhhh}^{(2)}\big|^\mathrm{gen.}_\text{THDM}  +  \Gamma_{hhhh}^{(2)}\big|^\mathrm{subloop}_\text{THDM} -  \Gamma_{hhhh}^{(2)}\big|^\mathrm{gen.}_\text{SM} -  \Gamma_{hhhh}^{(2)}\big|^\mathrm{subloop}_\text{SM} \right] =  \nonumber\\
    & =
        \frac{1}{(4\pi)^4}\frac{3}{2} (c_\beta h_t-s_\beta h_t^\prime)^2(s_\beta h_t+ c_\beta h_t^\prime)^4 \left(2\pi^2-7+14\llog M_{H^+}^2 + 6 \llog^2 M_{H^+}^2\right)\,,
\end{align}
which agrees with Eq.~(52) in \ccite{Bahl:2020jaq}. This result can easily be generalised to the complex THDM by replacing the two brackets containing the Yukawa couplings with absolute values.


\subsection{Cross-checks in the Next-to-Minimal Supersymmetric Standard Model}

In this Section, we discuss the performance potential of the framework presented in \cref{sec:generic_calc,sec:FA_mapping} and also provide further verification by applying it to a Supersymmetric (SUSY) model. SUSY models usually feature large particle spectra, many different mass scales, and a large set of complex Feynman rules. Therefore, tensor and integral reductions at the level of model-specific particle insertions often involve very large expressions and are increasingly difficult with standard tool chains. In such cases, one can largely benefit from the algorithm presented in this work because \textit{(i)} the number of Feynman diagrams can be significantly reduced using symmetry relations and \textit{(ii)} the tensor and integral reductions have already been carried out in a previous step, at the level of generic fields. 

In order to demonstrate the case made above and to further cross-check the correctness of our generic expressions and their mapping to diagrams generated via \FeynArts, we consider in this Section the two-loop corrections of order $\mathcal{O}(\alpha_t^2)$ to the trilinear Higgs self-coupling in the charge-parity (\cp) violating version of the Next-to-Minimal Supersymmetric Standard Model (NMSSM), which was already studied in \ccite{Borschensky:2022pfc} and implemented in the computer program \texttt{NMSSMCALC} \cite{Baglio:2013iia}. For the cross-check with \ccite{Borschensky:2022pfc}, we compare our results numerically on a diagram-by-diagram level with the implementation in \texttt{NMSSMCALC} and find full agreement. Furthermore, we are able to demonstrate the performance improvement in our approach by comparing the number of Feynman diagrams before and after symmetry relations have been used --- as shown in \cref{tab:NMSSMsymmetryfacs}. For the counting, we create diagrams at the \texttt{Classes} level (in \FeynArts), \ie we consider different field types but we do not expand generation and colour indices. In the \cp-violating case, we cannot identify diagrams that would be identical if internal fermion propagators were reversed. Therefore, we distinguish between the \cp-conserving (numbers in brackets) and \cp-violating cases in \cref{tab:NMSSMsymmetryfacs}. We find a reduction in the number of potentially non-vanishing diagrams of roughly one order of magnitude for both the genuine two-loop diagrams and the diagrams with counterterm insertions. In addition, we find full numerical agreement with the calculation presented in \ccite{Borschensky:2022pfc} providing an extensive independent cross-check of our results.

\begin{table}[tb]
    \centering
    \begin{tabular}{c|c|c}
       diagrams        &  topology-level & field-level \\ \hline
      genuine two-loop & $39  \to 12$ & $213    \to  67(32)$ \\
      subloop         & $15  \to 5$  & $36    \to 12(7)$ 
    \end{tabular}
    \caption{Reduction of the number of two-loop 
    $\mathcal{O}(\alpha_t^2)$ three-point diagrams in the CP-violating(-conserving) NMSSM, resulting from the application of the algorithm described in \cref{sec:diagram_reduction}. The counting ignores generation indices of stops/sbottoms, electroweakinos, and Higgs bosons that appear on the internal lines.}
    \label{tab:NMSSMsymmetryfacs}
\end{table}

It has been shown in the past, that higher-order corrections to the SM-like trilinear Higgs self-coupling in SUSY models can play a crucial role in a precise description of \eg double Higgs production and Higgs-to-Higgs decays \cite{Brucherseifer:2013qva,Nhung:2013lpa,Muhlleitner:2015dua,Borschensky:2022pfc}. Therefore, the results derived in this work are also of particular interest for SUSY models. 

\section{Two-loop corrections to the trilinear Higgs coupling in the general singlet extension}
\label{sec:applications}
\label{sec:SSM}

In the following, we discuss a new calculation that has been performed by applying the generic results presented in \cref{sec:generic_calc} to the general (\ie non-$\mathbb{Z}_2$-symmetric) singlet extension of the SM, using the methods described in \cref{sec:FA_mapping}.

The calculation of the trilinear Higgs self-coupling in the SM extended with a real singlet so far has been performed in full generality only at the full one-loop level, see \eg \ccite{Bahl:2023eau}. At the two-loop order, only the leading corrections in a simpler version featuring a $\Ztwo$-symmetry are known, see \ccite{Braathen:2019zoh} and \cref{sec:SSMZ2}. In this Section, we present the calculation of the corresponding two-loop corrections in the general real-singlet extension. The potential for this model is written in terms of $\Ztwo$-conserving terms $V_{\Ztwo}(\Phi,S)$, \cf \cref{eq:SSMZ2:pot}, and a $\Ztwo$-breaking part 
\begin{equation}
    V_{\text{SSM}}(\Phi,S) = V_{\Ztwo}(\Phi,S) + 
    \frac{\kappa_S}{3} S^3 +
    \kappa_{SH} S|\Phi|^2
    \,+\,\tilde{t}_S S\,.
    \label{eq:SSM:pot}
\end{equation}
The tadpole term $\tilde{t}_S S$ can be eliminated with the use of appropriate field- and coupling-redefinitions~\cite{Espinosa:2011ax}. 
In the general case, both the singlet and the doublet receive VEVs, and can be expanded as
\begin{equation}
    S = s + v_S \qquad\text{and}\qquad
    \Phi = \frac{1}{\sqrt{2}} \begin{pmatrix} \sqrt{2}G^+ \\ v+h+i G \end{pmatrix}.
\end{equation}
The $\Ztwo$-breaking allows for $s$ and $h$ to mix into the mass eigenstates $h_1$ and $h_2$ via the mixing angle $\alpha$. We assume that the SM-like Higgs boson is the lighter scalar state and that there is a sizeable mass splitting, \ie
\begin{equation}
m_{h_1}\simeq
\unit[125]{GeV}
\ll
m_{h_2}\,.
\label{eq:SSM:mh}
\end{equation}
In the following, we first discuss the calculation in the \MS scheme and then generalise the one-loop OS-like prescription for the renormalisation of $\lambda_{h_1 h_1 h_1}$ described in Appendix~C.2 of \ccite{Bahl:2023eau}, to the two-loop level. A detailed description of the notation used in the following can be found therein.
We consider the case of vanishing tree-level mixing between $h_1$ and $h_2$. To simplify the two-loop calculation, and following \cref{eq:SSM:mh}, we also set $m_{h_1}=0$, which in turn implies $\lambda_H=0$ (see the discussion in \cref{sec:SSMZ2}). To further simplify the notation, we re-label the singlet-like state $h_2\equiv s$ and the doublet-like state $h_1\equiv h$.

\begin{figure}[tb]
    \centering
    \includegraphics[width=0.9\textwidth]{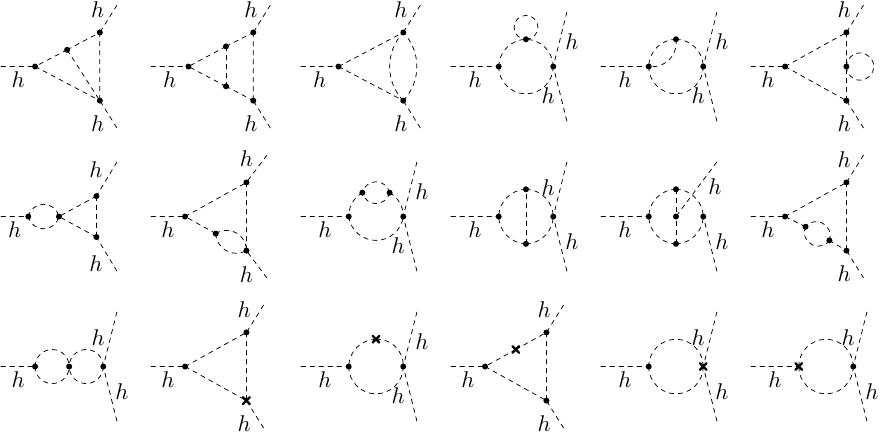}
    \caption{All BSM-type diagrams contributing to the SM-like trilinear Higgs self-coupling in the general SSM using the approximations described in the text. Internal scalars can be $h$ or $s$ whereas diagrams with $hhh$- and $hhhh$-couplings vanish due to the approximations we have applied.}
    \label{fig:SSMdiags}
\end{figure}

\paragraph{\underline{Calculation in the \MS scheme}:}
In order to be able to generate Feynman diagrams for the SSM using \FeynArts, we implemented the SSM in \texttt{SARAH}~\cite{Staub:2008uz,Staub:2009bi,Staub:2010jh,Staub:2012pb,Staub:2013tta} and used its \FeynArts~output to create a model file containing all Feynman rules for tree-level and one-loop counterterm vertices.
All two-loop three-point diagrams required for the calculation (after using the symmetry relations described in \cref{sec:diagram_reduction}) are shown in \cref{fig:SSMdiags} and have been processed automatically using the procedures described in the previous chapters. We are left with the renormalisation of the trilinear coupling itself.

Applying the two assumptions, $m_{h}=0$ and $\alpha=0$, \textit{after} performing the renormalisation transformation leads to the following two-loop vertex counterterm in the \MS scheme:
\begin{align}
    \delta^{(2)}_\text{CT} \lambda_{hhh}^{\MS} = {}& 3 v \, \delta^{(2)}_\text{CT}\lambda_H^{\MS} 
    -\frac{9\, v}{2}\delta^{(1)}_{\rm CT}\lambda^{\MS}_H\delta_{\rm CT}^{(1)}Z_{hh}
    \nn \\ 
    & +\frac{3}{4 v_S}\delta_{\rm CT}^{(1)}Z_{sh}\left( v\,\kappa_{SH}\delta_{\rm CT}^{(1)}Z_{sh} -2(v_S\,\delta_{\rm CT}^{(1)}\kappa_{SH} + v_S^2\delta_{\rm CT}^{(1)}\lambda_{SH})\right)\,,
    \label{eq:SSM:ctMS}
\end{align}
where $\delta_{\rm CT}^{(n)}x^{\MS}$ is the $n$-loop counterterm of the parameter $x$ defined in the \MS scheme.
The one-loop counterterm of the doublet VEV, $\delta^{(1)}_\text{CT}v$, does not receive any BSM corrections in the limit $\alpha=0$ and has therefore already been set to zero in \cref{eq:SSM:ctMS} --- in which we only include terms that contribute to the leading two-loop BSM corrections --- while its two-loop counterterm $\delta^{(2)}_\text{CT}v$ does not appear either as $m_h$ has been set to zero.
All \MS counterterms that enter the renormalised $\hat{\lambda}_{hhh}$ at the two-loop level, either as genuine two-loop counterterms or via subloop renormalisation, can be extracted from the renormalisation group equations (RGEs) computed with {\tt SARAH}.

We have successfully checked that the \MS counterterms render the scalar one-, two- and three-point functions UV-finite at the one- and two-loop levels. Their algebraic expansions in terms of $d=4-2\epsilon$ are given in \cref{app:SSM}. Furthermore, we have verified that the evanescent UV-finite terms, generated by $\order{\epsilon}$-terms of one-loop functions multiplied with $\order{\epsilon^{-1}}$ poles, cancel exactly. Moreover, the field renormalisation constants $\delta_{\rm CT}^{(1)}Z_{ij}$, which are defined by the relation between the bare and the renormalised fields
\begin{equation}
    \begin{pmatrix}
    h \\ s
    \end{pmatrix}
    \to
    \left[
    \mathbb{1}
    +
    \frac{1}{2}
    \begin{pmatrix}
     \delta^{(1)}_\text{CT}Z_{hh} & \delta^{(1)}_\text{CT}{Z}_{sh}\\
     \delta^{(1)}_\text{CT}{Z}_{hs} & \delta^{(1)}_\text{CT}Z_{ss}\\
    \end{pmatrix}
    \right]
    \begin{pmatrix}
    h \\ s
    \end{pmatrix}
    \,,
    \label{SSM:fieldnorm}
\end{equation}%
drop out in the final result (after including LSZ factors, see the discussion below) and thus need not be specified further.

The complete result for the renormalised trilinear Higgs self-coupling in a fully \MS scheme (with tadpoles treated in the tadpole-free scheme, see \eg~\ccite{Martin:2014cxa,Martin:2015lxa,Martin:2015rea,Martin:2016xsp,Martin:2018yow} and references therein) is 
\begin{align}
    (4\pi)^4\delta^{(2)}\lambda_{hhh}\big|^{\MS} = &
        -\frac{3}{8}\frac{\kapSH^2 v}{v_S^5} \Big[
          6 \kapS v_S^2 (3 \lnMS+\lnMSs-3)
         +8 \kapSH v_S^2 (-2 \lnMS+\lnMSs+1)
       \nn \\ & \qquad\qquad    +  \kapSH v^2 (-23 \lnMS-3 \lnMSs+35)
      \Big] \nn \\
    &
    -\frac{1}{m_s^2}\frac{3 \kapSH^2 v}{16 v_S^6} \Big[
            \kapSH^2  v^4 (35-17 \lnMS)
          -4 \kapS^2  v_S^4 (\lnMS-1) \nn
         \\ & \qquad\qquad+4 \kapSH v_S^2 v^2 \left(\kapS (3 \lnMS-8) - 6 \kapSH (\lnMS-1)\right)\nn
    \Big] \\
    &
    +\frac{1}{m_s^4}\frac{\kapSH^3 v^3 (\lnMS-2)}{16 v_S^7} \Big[
          4 \kapS^2 v_S^4 
         +4 \kapSH v_S^2 v^2 (2 \kapSH-3 \kapS)
         +9 \kapSH^2 v^4
    \Big] \nn \\
    &
    +m_s^2 \frac{9 \kapSH^2 v}{4 v_S^4} \left(4 \lnMS+\lnMSs-4\right)\,.
    \label{eq:SSM:lammhhh2LMS}
\end{align}
In order to deal with proper on-shell external states and include corrections to external legs, we dress the relevant lower-order renormalised couplings with LSZ factors:
\begin{align}
\delta^{(2)}_\text{LSZ} \lambda_{hhh}^{\MS} ={}&
-3\left[\frac{1}{2} Z_{hh}^{(1)}\cdot\hat\lambda_{hhh}^{(1)}\big|^\MS + Z_{hs}^{(1)}\cdot \hat\lambda_{hhs}^{(1)}\big|^\MS + \left(Z_{hs}^{(1)}\right)^2\cdot\lambda_{hss}^{(0)}\big|^\MS \right]\,,
\end{align}
where $\hat\lambda_{ijk}^{(1)}\big|^\MS$ and $\lambda_{ijk}^{(0)}$ are respectively the renormalised one-loop and the tree-level vertices in the \MS scheme, and
\begin{subequations}
    \label{eq:SSM:LSZ}
    \begin{align}
        Z_{hh}^{(1)} & = \Sigma_{hh}^{\prime}(0) + \delta_{\text{CT}}^{(1)} Z_{hh}^{(1)}\,,\\
        Z_{ss}^{(1)} & = \Sigma_{ss}^{\prime}(0) + \delta_{\text{CT}}^{(1)} Z_{ss}^{(1)}\,,\\
        Z_{hs}^{(1)} & = -\left[\frac{\Sigma_{hs}(p^2) - \delta_{\rm CT}^{(1)}m_{hs}^2+ \frac{m_s^2}{2}\delta_{\text{CT}}^{(1)}{Z}_{hs}}{p^2-m_s^2} \right]_{p^2=m_h^2=0}\,,\\
        Z_{sh}^{(1)} & = -\left[\frac{\Sigma_{sh}(p^2) - \delta_{\rm CT}^{(1)}m_{hs}^2 - \frac{m_s^2}{2}\delta_{\text{CT}}^{(1)}{Z}_{sh}}{p^2-m_h^2} \right]_{\substack{p^2=m_s^2\\ m_h^2=0 \\ \phantom{|}}}\,.
    \end{align}
\end{subequations}
Analytic results for the LSZ corrections are given in \cref{app:SSM}. The mass mixing counterterm in \cref{eq:SSM:LSZ}, 
\begin{equation}
\delta_{\rm CT}^{(1)}(m_{hs}^2)^\MS =v\,\delta_{\rm CT}^{(1)}\kappa_{SH}^{\MS} + v\,v_S\,\delta_{\rm CT}^{(1)}\lambda_{SH}^{\MS}\,,
\end{equation}
consists of the one-loop $\MS$ counterterms for $\kappa_{SH}$ and $\lambda_{SH}$, which renders the off-diagonal part of the renormalised self-energy UV-finite, \cf \cref{eq:app:SSM:ZhsMS}. 

From \cref{eq:SSM:lammhhh2LMS} one can derive the leading correction in the limit of a heavy singlet mass compared to all other mass parameters,
\begin{equation}
\delta^{(2)}\lambda_{hhh}\big|^{\MS}\approx\frac{9 v m_{s}^2 \kappa_{SH}^2 }{4(4\pi)^4v_S^4}\left(\overline{\ln}^2m_{s}^2 + 4 \overline{\ln}m_{s}^2 - 4\right),
\label{eq:SSM:MS}
\end{equation}
which shows that, in this parametrisation, ensuring a proportionality relation of the form $v_S\propto m_s$ is required in order to properly decouple the singlet in the limit $m_s\to\infty$. Furthermore, \cref{eq:SSM:MS} can have a rather large renormalisation scale dependence since the $\overline{\ln}$-terms can receive a strong enhancement when the Lagrangian trilinear coupling $\kappa_{SH}$ is large and $v_S\ll m_{s}$. In the following, we will show that this can be avoided by using an OS renormalisation prescription that removes the dependence on the renormalisation scale and ensures manifest decoupling as well.

\paragraph{\underline{Calculation in the OS scheme}:} 
Analogously to the \MS scheme, we start with the vertex counterterm $\delta^{(2)}_\text{CT} \lambda_{hhh}^{\text{OS}}$. In the OS scheme, we rewrite the trilinear self-coupling in terms of the input parameters:
\begin{equation}
    \underbrace{m_h,\, m_s,\, \alpha,\, t_s,\,t_h,v}_{\text{on-shell}},\, \underbrace{v_S,\kappa_S,\,\kappa_{SH}}_{\MS},\,
\end{equation}
where $t_h$ and $t_s$ are the tadpoles of the doublet-like and singlet-like fields, respectively. In the general singlet extension, there are not enough mass parameters and mixing angles to trade also the singlet VEV $v_S$ and the trilinear couplings $\kappa_S$ and $\kappa_{SH}$ for an OS mass or mixing angle. One way to renormalise them OS would be by connecting them to certain decay observables, see \eg~\ccite{Denner:2018opp}. In the present work, we limit ourselves to an \MS renormalisation of these three parameters.

Within this parametrisation, we renormalise all input parameters and the fields by $x\to x+\delta^{(1)}_\text{CT}x+\delta^{(2)}_\text{CT}x$ and set $\alpha=0$, $m_h=0$ \textit{after performing the renormalisation transformation}.
When renormalising the mixing angle $\alpha$, which enters in the scalar mixing matrix, the field renormalisation transformation also receives an additional contribution from the mixing angle counterterm \cite{Kanemura:2004mg}. Thus, we replace the off-diagonal field renormalisation factors in \cref{SSM:fieldnorm} by
\begin{equation}
    \delta^{(1)}_{\text{CT}} Z_{sh/hs} \to \delta^{(1)}_{\text{CT}} \hat{Z}_{sh/hs}=\delta^{(1)}_{\text{CT}} Z_{sh/hs}\pm 2\,\delta_{\text{CT}}^{(1)}\alpha
    \label{eq:SSM:Zhshift}
\end{equation}
Taking all counterterms up to two-loop order into account, we arrive at
\begin{align}
    \delta^{(2)}_\text{CT} \lambda_{hhh}^{\text{OS}} ={}& 
    \frac{3}{v^2}\left[
          \delta^{(2)}_\text{CT}t_{h} 
        - v\,\delta^{(2)}_\text{CT}m_{h}^2
        - \left(\delta^{(1)}_\text{CT}\alpha\right)^2 m_s^2
        + \frac{3}{2}\left(
             \delta^{(1)}_\text{CT}t_{h}
             -v\, \delta^{(1)}_\text{CT}m_{h}^2 
           \right)\delta^{(1)}_\text{CT}Z_{hh}
    \right] \nonumber\\
    & \qquad-
   \delta^{(1)}_\text{CT}\hat{Z}_{hs} \left[\frac{v^3\kappa_{SH}}{4 v_s}\delta^{(1)}_\text{CT}\hat{Z}_{hs}
    +\frac{v\, m_s^2}{2}\delta^{(1)}_\text{CT}\alpha 
    \right]\,.
    \label{eq:SSM:ctOS}
\end{align}
Similar to the \MS scheme, we consider the external-leg corrections
\begin{align}
\label{eq:SSM:LSZOS}
\delta^{(2)}_\text{LSZ} \lambda_{hhh}^{\text{OS}} ={}&
-3\left[\frac{1}{2} Z_{hh}^{(1)}\cdot\lambda_{hhh}^{(1)}\big|^\text{OS} + Z_{hs}^{(1)}\cdot \lambda_{hhs}^{(1)}\big|^\text{OS} + \left(Z_{hs}^{(1)}\right)^2\cdot\lambda_{hss}^{(0)}\big|^\text{OS} \right]\,.
\end{align}
Due to the OS renormalisation of $\alpha$, the mixing on the external legs is already accounted for, \ie $Z_{hs}^{(1)}=\delta_{\rm CT}^{(1)}Z_{hs}/2$. Therefore, the off-diagonal contributions in \cref{eq:SSM:LSZOS} only contain wave-function renormalisation factors, which cancel those appearing in \cref{eq:SSM:ctOS} and in \cref{fig:SSMdiags}, and therefore do not contribute to the final result.
It is interesting to note that, although we have assumed $\alpha=0$, the counterterm of the mixing angle, $\delta^{(1)}_\text{CT} \alpha$, still contributes to the finite part of the final result unlike what is the case at the one-loop order. Two-loop counterterms for $\alpha$ and the fields are not required in the scenario of vanishing tree-level mixing. 

The two-loop tadpole counterterm entering \cref{eq:SSM:ctOS} reads:
\begin{equation}
\label{eq:SSM:dtadh2L}
    \delta^{(2)}_\text{CT}t_{h} = - t^{(2)}_{h}
                        - \frac{1}{2}\delta^{(1)}_\text{CT}Z_{hh} \delta^{(1)}_\text{CT}t_{h}
                        -  \delta^{(1)}_\text{CT}t_{s} \left(\frac{1}{2}\delta^{(1)}_\text{CT}\hat{Z}_{sh} + \delta^{(1)}_\text{CT}\alpha\right)
                        ,\,
\end{equation}
where $\delta^{(1)}_\text{CT}t_{h/s} = -t^{(1)}_{h/s}$ are the one-loop tadpole counterterms and $t_{h/s}^{(i)}$ are the $i$-loop one-point functions.
At each loop order, the mass counterterms are fixed by requiring a vanishing on-shell renormalised self-energy.
The two-loop Higgs mass counterterm in the OS scheme is thus given by
\begin{align}
    \delta^{(2)}_\text{CT}m_{h}^2 &= \left[\Sigma^{(2)}_{hh}(p^2) 
    -\hat{\Sigma}^{(1)}_{hh}(p^2)\cdot \frac{\partial }{\partial p^2}\hat{\Sigma}^{(1)}_{hh}(p^2)\right]_{p^2=m_h^2=0}\nonumber \\
    &\quad - \delta^{(1)}_\text{CT}Z_{hh} \delta^{(1)}_\text{CT}m^2_{h}
     - \frac{m_s^2}{4}\left(\delta_{\text{CT}}^{(1)}\hat{Z}_{hs}\right)^2
     - m_s^2\left(Z_{hs}^{(1)}\right)^2
     \,,
    \label{eq:SSM:mhct}
\end{align}
where $\hat{\Sigma}^{(1)}_{hh}(p^2) \equiv \Sigma_{hh}(p^2)-\delta_{\text{CT}}^{(1)}m_h^2+(p^2-m_h^2)\delta_{\text{CT}}^{(1)}Z_{hh}$ is the diagonal part of the renormalised one-loop Higgs boson self-energy. Note that, unlike in the \MS scheme, the off-diagonal mass counterterm vanishes when renormalising $\alpha$ in the OS scheme and therefore does not contribute here. The counterterm of $\alpha$ can be fixed by demanding vanishing mixing\footnote{Alternative choices can be motivated by considering the decay $s\to h h $, which is forbidden at tree-level but loop-induced if $\alpha=0$. We have checked that extracting $\delta^{(1)}_{\text{CT}}\alpha$ for instance from the condition $\hat\lambda_{hhs}^{(1)}\big|^{\text{OS}}\stackrel{!}{ =} 0$ is also a viable option that yields very similar results as the ones presented here.}, \ie 
\begin{equation}
    \delta^{(1)}_{\text{CT}}\alpha = -\frac{\Sigma_{hs}^{(1)}(0)}{m_s^2}\,.
\end{equation}
The two remaining parameters, which are to be renormalised at one loop in order to obtain a UV-finite result, are $\kappa_S$ and $\kappa_{SH}$. We choose to define them in the $\MS$ scheme. For their corresponding counterterms we refer to the derivation of the one-loop renormalisation in the \MS scheme in \cref{app:SSM}.

With these ingredients, we obtain a UV-finite result for the renormalised trilinear Higgs self-coupling in the OS scheme. As a further cross-check of our expressions, we have verified that all field normalisation factors, $\delta_{\text{CT}}^{(1)}Z_{ij}$ defined in \cref{SSM:fieldnorm}, cancel out in the sum of all contributions.
The final two-loop shift to the renormalised coupling is rather simple and reads
\begin{align}
    (4\pi)^4\delta^{(2)}\lambda_{hhh}\big|^{\mathrm{OS}}  &=  (4\pi)^4\left(\delta^{(2)}_\text{\cref{fig:SSMdiags}} \lambda_{hhh}^{\text{OS}} +\delta^{(2)}_\text{CT} \lambda_{hhh}^{\text{OS}} +\delta^{(2)}_\text{LSZ} \lambda_{hhh}^{\text{OS}} \right)\nn \\
    & = 
    -\frac{9\, \kappa_{SH}^3\, v^3}{2\, v_S^5}\nn\\ &\quad \,
    -\frac{3\, \kappa_{SH}^3\, v^3}{2\, m_s^2\, v_S^4}\left[ \left(\kappa_{S}+2\, \kappa_{SH}\right) \overline{\ln}m_s^2 
     - 2 \left(\kappa_{S}-\kappa_{SH}\right) -
      3\,\kappa_{SH} \frac{v^2}{v_S^2} \right] \nn \\
      & \quad\,- \frac{\kappa_{SH}^3\, v^3}{8\, m_s^4\, v_S^3}\left[
      4\,\kappa_S^2 + \kappa_{SH}\left(\frac{7}{2}\kappa_{SH}-12\,\kappa_S\right)\frac{v^2}{v_S^2}
      +9\,\kappa_{SH}^2\frac{v^4}{v_S^4}
      \right]\,,
      \label{eq:SSM:twoloopresultOS}
\end{align}
which can be safely evaluated for $m_s\to\infty$. We note that the tree-level expression for $\lambda_{hhh}$ in the limit $\alpha=0$ is identical to the SM and therefore vanishes in the applied approximation of $m_h=0$. Taking $m_s$ very large, the one-loop corrections also vanish (\cf \cref{eq:SSM:oneloopresult}) and we obtain a rather simple result for the full two-loop renormalised trilinear coupling: 
\begin{equation}
    \hat{\lambda}^{(2)}_{hhh}\big|^{\mathrm{OS}}(\alpha=0,m_s,\kappa_S,\kappa_{SH}) \overset{m_s\gg m_h}{\approx} \frac{3 m_h^2}{v}-\frac{1}{(4\pi)^4}\frac{9 \kappa_{SH}^3 v^3}{2 v_S^5}\,.
\end{equation}
It is important to remark that --- despite the OS renormalisation of the masses, mixing angle, and fields --- the two-loop shift in \cref{eq:SSM:twoloopresultOS} has a left-over dependence on the renormalisation scale $Q$, via terms of the form $\propto m_s^{-2}(\overline{\ln}m_s^2)$, which is suppressed by the squared singlet mass. This is due to the parameters $\kappa_S$ and $\kappa_{SH}$ being renormalised in the $\MS$ scheme. In particular, $\kappa_{SH}$ enters the calculation of $\hat\lambda_{hhh}$ for the first time at the one-loop order (see \cref{eq:SSM:oneloopresult}) and, therefore, its running induces a two-loop effect. We have explicitly checked that the two-loop shift induced via a leading-log running of $\kappa_{SH}$ in the one-loop result is cancelled by the logarithmic dependence of \cref{eq:SSM:twoloopresultOS} on $m_s$, \ie that
\begin{equation}
 \frac{\partial}{\partial \overline{\ln}m_s^2 } \delta^{(2)} \lambda_{hhh}\big|^{\mathrm{OS}} = -\left( \frac{\partial}{\partial \kappa_{SH}}\delta^{(1)}\lambda_{hhh}\big|^\text{OS}\right)\cdot\frac{\partial}{\partial \overline{\ln}m_s^2 } \left( \left. \delta^{(1)}_\text{CT}\kappa_{SH}^{\MS}\right|_{\frac{1}{\epsilon}\to-\overline{\ln}{m_s^2}}\right)
\label{eq:SSM:llcancellation}
\end{equation}
which is a further strong cross-check of our result.

\paragraph{\underline{The $\Ztwo$ limit}:}
Additional cross-checks of our results can be performed by taking the $\Ztwo$-conserving limit and comparing with the results in the literature and \cref{sec:SSMZ2}. When computing a quantity $\mathcal{M}$ in within the general SSM, the $\Ztwo$ limit is given by
\begin{equation}
    \mathcal{M}^{\text{\zssm}} = \lim_{v_S\to0} \left\{ \mathcal{M}^{\text{SSM}}\left(
    \lambda_{SH}\to c_1,\,
    m_s^2 \to c_2
    \right)\right\}
\end{equation}
with
\begin{equation}
    c_1 = \frac{2(m_s^2-\mu_S^2)}{v^2}, \quad\text{and}\quad
    c_2 =\mu_S^2 + 2 \kappa_S v_S + 6 \lambda_S v_S^2 + \lambda_{SH} v^2/2\,.
\end{equation}
In this limit, all two-loop one-, two- and three-point functions computed in \cref{sec:SSMZ2} have been reproduced.

\paragraph{\underline{Example numerical results}:}
Finally, we present a numerical example for $\kappa_\lambda$. We choose to set $\kappa_{SH} = -10\cdot v_s = - 3 \tev$. If not stated otherwise, $\kappa_S = \kappa_{SH}$ is chosen and the input renormalisation scale of $\kappa_S$ and $\kappa_{SH}$ is fixed to $m_s$.

\begin{figure}
    \centering \includegraphics[width=\textwidth]{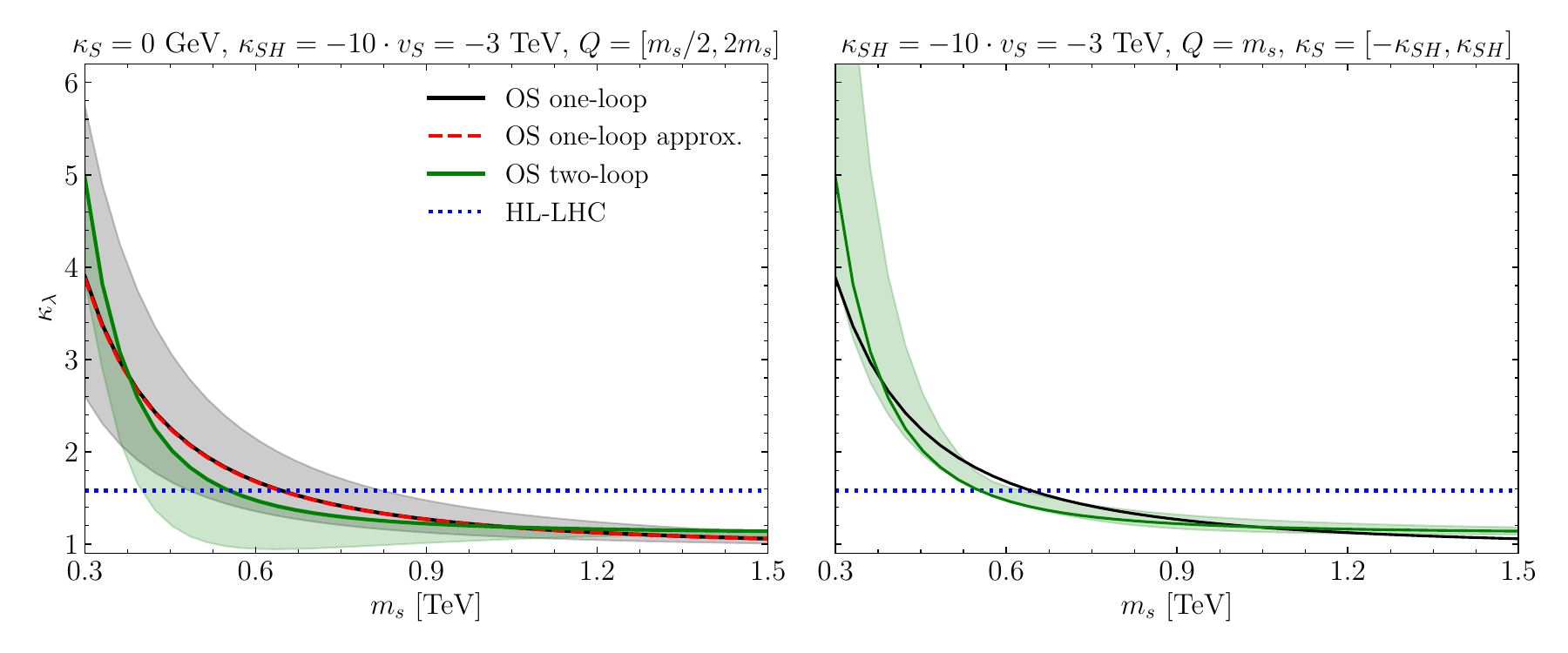}
    \caption{$\kappa_\lambda$ as a function of $m_s$ computed in the general singlet extension of the SM. \textit{Left:} Comparison of the full one-loop (black solid curve), approximated one-loop (red dashed curve), and two-loop results (green solid curve) and their scale dependence. The blue-dashed line corresponds to the $2\sigma$ interval of the HL-LHC projection for $\kappa_\lambda$ from \ccite{CMS:2025hfp}. \textit{Right:} Dependence of the two-loop result on $\kappa_S$.}
    \label{fig:SSM_OS}
\end{figure}

In the left panel of \cref{fig:SSM_OS}, $\kappa_\lambda$ is shown as a function of $m_s$. The differences between the full one-loop (black solid, computed using \anyH~\cite{Bahl:2023eau}) and the approximated one-loop (red dashed, setting $m_{h} = 0$ in the one-loop corrections, \cf \cref{eq:SSM:oneloopresult}) results are negligibly small, thus validating the approximation used for the calculation of the leading two-loop corrections (green solid). We note that at the tree level we always use $m_h=125\gev$ since $m_h=0$ would introduce a rather large error. The two-loop corrections enhance $\kappa_\lambda$ by $\sim 1.2$ for $m_s \sim 300\gev$ compared to the one-loop result. The shift induced by the two-loop corrections shrinks down for rising $m_s$ and then increases again for $m_s\gtrsim 800\gev$. The reason why the asymptotic behaviour of the two-loop curve is different from the one-loop curve is that the singlet-specific one-loop corrections vanish for large $m_s$ while the corresponding two-loop corrections contain a part that is independent of $m_s$ but only suppressed by $v_S$, \cf the first term in \cref{eq:SSM:twoloopresultOS}. However, for a proper decoupling a relation of the form $v_S\propto m_s$  (or alternatively $\kappa_S=\kappa_{SH}=0$) is required, which induces the appropriate asymptotic behaviour.

The left panel of \cref{fig:SSM_OS} also shows the scale dependence of the one- and two-loop corrections induced by defining $\kappa_{SH}$ in the \MS scheme (while all other parameters entering the one-loop correction are fixed in the OS scheme). The coloured bands are derived by varying the renormalisation scale between $Q=m_s/2$ and $Q=2 m_s$, and using the leading-log approximation to relate $\kappa_{SH}(m_S)$ to $\kappa_{SH}(Q)$.\footnote{Note that we also run $\kappa_{S}$ from $m_S$ to $Q$, using the leading-log approximation and starting at $m_S$ with the input $\kappa_S(m_S)=\kappa_{SH}(m_S)$. Thus, in general we have $\kappa_{S}(Q)\neq\kappa_{SH}(Q)$.} Although the scale dependence is very sizeable at the one-loop level, this dependence is significantly reduced when including the two-loop corrections, clearly showing their numerical importance --- \cf the discussion around \cref{eq:SSM:llcancellation}.

Finally, in the right panel of \cref{fig:SSM_OS}, we investigate the dependence of the two-loop corrections on $\kappa_S$, which first enters at the two-loop level. We assess this dependence by varying $\kappa_S$ between $-\kappa_{SH}$ and $\kappa_{SH}$. We observe that the dependence on $\kappa_S$ is sizeable and especially pronounced for low $m_s$, while it shrinks down for higher $m_s$. However, even for high $m_s$, it remains of the same size as the difference between the one- and two-loop predictions for $\kappa_\lambda$ computed for $\kappa_S = 0$. This illustrates the importance of this new class of two-loop corrections, which we have computed for the first time in this work.


\section{Conclusions}
\label{sec:conclusions}

The investigation of the Higgs potential is one of the main goals of the LHC programme and of future colliders. High-precision theoretical predictions are crucial to exploit the experimental constraints, or (in the future) measurements, on the trilinear and quartic Higgs couplings, in particular in order to understand implications for concrete BSM scenarios. As shown in the literature and also confirmed by our new two-loop results, one-loop predictions for the trilinear and quartic Higgs couplings are in this regard not sufficient for an accurate result.

With this motivation in mind, we have presented a generic two-loop calculation of the scalar $n$-point functions (with $n\leq 4$) with identical external particles in the zero-momentum approximation. Besides the genuine two-loop contributions, we have also presented results for the occurring subloop-renormalisation contributions. Together, these constitute the necessary ingredients for calculations of the effective trilinear and quartic self-couplings of the detected Higgs boson, corresponding to the operators that are (or will be, for the quartic coupling) constrained by LHC data and can also be understood as operators in the Higgs Effective Field Theory (HEFT)~\cite{Feruglio:1992wf,Burgess:1999ha,Giudice:2007fh,Grinstein:2007iv,Alonso:2012px,Azatov:2012bz,Buchalla:2012qq,Contino:2013kra,Buchalla:2013rka,Alonso:2014rga,Guo:2015isa,Buchalla:2015qju,Buchalla:2017jlu,Buchalla:2018yce,Falkowski:2019tft,Cohen:2020xca}. 

Building upon a unique canonical form of Feynman diagrams, we have demonstrated how the number of diagrams to compute can be reduced using symmetry relations. Moreover, we have worked out an on-the-fly reduction of the appearing loop integrals allowing for a stable numerical evaluation. These evaluation routines are available publicly in the form of the new \texttt{Python} package \texttt{Tintegrals}. Lastly, we have shown how our generic results can be mapped to amplitudes generated via \FeynArts, thereby facilitating the application of our results to a wide range of theories. 

Using these routines, we have performed a series of cross-checks against existing results in the literature. Moreover, we have presented new two-loop results for the trilinear Higgs coupling in the general scalar singlet extension of the SM --- these are in particular the first two-loop results for $\lambda_{hhh}$ including the effect of Lagrangian trilinear couplings. Our numerical results show that the projected HL-LHC bounds on $\kappa_\lambda$ can probe large parts of the singlet-extended SM and that the newly calculated two-loop corrections significantly reduce the theoretical uncertainties.

The obtained generic results are a core ingredient needed for automatic two-loop predictions of the trilinear and quartic Higgs couplings in arbitrary BSM models (\eg~in the \texttt{anyH3} framework), which we leave for future endeavours. Our work also offers a base to build on to investigate high-precision predictions of Higgs (self-)couplings with non-vanishing external momenta, and to tackle the Goldstone Boson Catastrophe~\cite{Elias-Miro:2014pca,Martin:2014bca,Kumar:2016ltb,Braathen:2016cqe,Braathen:2017izn,Goodsell:2019zfs} in this context. 


\section*{Acknowledgements}
\sloppy{
We thank Mark Goodsell for organising the \textit{Workshop on Automatic Phenomenology}, where this work was initiated. Moreover, we thank Felix Egle and Georg Weiglein for useful discussions. J.B.\ and M.G. acknowledge support by the Deutsche Forschungsgemeinschaft (DFG, German Research Foundation) under Germany's Excellence Strategy --- EXC 2121 ``Quantum Universe'' --- 390833306. This work has been partially funded by the Deutsche Forschungsgemeinschaft (DFG, German Research Foundation) --- 491245950. J.B.\ is supported by the DFG Emmy Noether Grant No.\ BR 6995/1-1.
}


\appendix


\section{Loop-corrected scalar couplings in the general singlet extendion of the Standard Model}
\label{app:SSM}

In this appendix, we renormalise the scalar sector of the general SSM in the approximation of vanishing tree-level mixing and of a heavy singlet mass. This is, in parts, required for the calculation of the two-loop shift presented in \cref{sec:SSM}.

\paragraph{\underline{Renormalisation in the \MS scheme:}} As already mentioned in the main text, \MS counterterms can in principle conveniently be derived from renormalisation group equations, which have been known for general renormalisable theories since a long time, see \eg \ccite{Machacek:1983tz,Machacek:1983fi,Machacek:1984zw,Sperling:2013xqa,Sperling:2013eva,Schienbein:2018fsw}, and are available for concrete models in \eg \texttt{SARAH} \cite{Staub:2010jh}. While this method can serve as a quite powerful cross-check of the UV-structure of the calculated diagrams (as we are introducing information from an independent calculation), it can have caveats in situations where approximations increase the number of scale-less integrals (as is the case in the calculation presented in this work) that we want to briefly comment on here. For demonstration purposes, we consider the scalar contributions to the one-loop beta function of $\kappa_S$. The result computed by \texttt{SARAH} consists of two contributions
\begin{equation}
    \beta^{(1)}_{\kappa_{S}} = 36\kappa_S\lambda_S + 6\kappa_{SH}\lambda_{SH}\,,
    \label{eq:betakaps}
\end{equation}
which can be understood by considering the UV-divergent diagrams of the one-loop scalar contributions to the $SSS$-vertex: 
\begin{equation}
    \delta^{(1)}\kappa_{S}|_{\text{UV}} \sim 36\kappa_S\lambda_S \textbf{B}_0(0,\mu_S^2,\mu_S^2)|_{\text{UV}} + 6\kappa_{SH}\lambda_{SH}\textbf{B}_0(0,\mu,\mu)|_{\text{UV}}
\end{equation}
the first term originates in the diagram with two internal singlet scalars. The second UV-divergent diagram has two doublet scalars in the loop. However, since $\mu^2=0$ (in our approximation) the second diagram is scale-less and therefore vanishes in dimensional regularisation. Thus, there is a mismatch between the beta function in \cref{eq:betakaps} and the counterterm $\delta^{(1)}_\text{CT}\kappa_S^{\MS}= \frac{18}{\epsilon}\kappa_S\lambda_S/(4\pi)^2$ for $\kappa_S$ relevant for the applied approximations.
Likewise, we find for the $\lambda_S$ counterterm $\delta^{(1)}_{\text{CT}} {\lambda_{S}}=18\lambda_S^2/(4\pi)^2$ instead of $\delta_{\text{CT}}^{(1)} {\lambda_{S}}=(18\lambda_S^2+1/2\lambda_{SH}^2)/(4\pi)^2$ even though we don't set $\lambda_{SH}$ to zero.
Having set scale-less integrals consistently to zero across the entire calculation, we arrive at the following \MS counterterms:
\begin{subequations}
\label{eq:app:SSM:MScts}
\begin{align}
    (4\pi)^2\delta^{(1)}_\text{CT}\kappa_S^{\MS}    & =  \frac{18}{\epsilon}\kappa_S\lambda_S\,,  \\
    (4\pi)^2\delta^{(1)}_\text{CT}\kappa_{SH}^{\MS} & = \frac{\lambda_{SH}}{\epsilon}\left(\kappa_{S}+2\kappa_{SH}\right)\,,\\
(4\pi)^2\delta^{(1)}_\text{CT}\lambda_{S}^{\MS}    & = \frac{18}{\epsilon}\lambda_S^2 \,,\\
(4\pi)^2\delta^{(1)}_\text{CT}\lambda_{H}^{\MS}    & = \frac{1}{2\epsilon}\lambda_{SH}^2 \,,\\
        (4\pi)^2\delta^{(1)}_\text{CT}\lambda_{SH}^{\MS}    & =  \frac{2\,\lambda_{SH}}{\epsilon}(3\lambda_S+\lambda_{SH})\,,  
\end{align}    
\end{subequations}
at the one-loop and
\begin{equation}
    \delta^{(2)}_\text{CT}\lambda_H^{\MS}= \frac{1}{(4\pi)^4}\left(\frac{1}{\epsilon^2} \lambda_{SH}^2(3\lambda_S+\lambda_{SH})
    - \frac{1}{2\epsilon} 3\lambda_{SH}^3\right)\,.
    \label{eq:app:SSM:dlam2LMSCT}
\end{equation}
at the two-loop order.

Using these counterterms, we can compute the lower-order renormalised couplings required for \cref{eq:SSM:LSZ}:
\begin{subequations}
\begin{align}
    \left.\hat\lambda^{(1)}_{hhh}\right|^{\MS} & = \frac{v\kappa_{SH}^2}{2\,v_S^3 (4\pi)^2}\left(\frac{v^2\kappa_{SH}}{m_s^2} - 3v_S\llog m_s^2\right)+ \mathcal{O}(m_h^2/m_s^2)\,, \label{eq:SSM:oneloopresultMS}\\
    \left.\hat\lambda^{(1)}_{hhs}\right|^{\MS} &=  \frac{\kappa_{SH}}{2\,v_S^3 (4\pi)^2}\left(v_S\llog m_s^2-\frac{v^2\kappa_{SH}}{m_s^2} \right)\left(3 m_s^2+\frac{3v^2\kappa_{SH}}{2 v_S}-v_S\kappa_S\right)+ \mathcal{O}(m_h^2/m_s^2)\,,\\
   \lambda^{(0)}_{hss} &= \frac{v}{v_s}\kappa_{SH}\,.
\end{align}
\end{subequations}
and the appropriate LSZ factors consisting of the renormalised \MS self-energies:
\begin{subequations}
\begin{align}
\label{eq:app:SSM:ZhsMS}
    \left.Z_{hh}^{(1)}\right|^{\MS} & =\frac{v^2\kappa_{SH}^2}{12v_S^2 m_s^2 (4\pi)^2} + \delta_{\rm CT}^{(1)}Z_{hh}\  \\
    \left.Z_{hs}^{(1)}\right|^{\MS} & = \frac{\hat{\Sigma}_{hs}(0)}{m_s^2}=\frac{1}{m_s^2}\left(\Sigma_{hs}(0) - v\,\delta_{\rm CT}^{(1)}\kappa_{SH}^\MS-v\,v_S\,\delta_{\rm CT}^{(1)}\lambda_{SH}^\MS - m_s^2\frac{\delta_{\rm CT}^{(1)}Z_{hs}}{2}\right)\\
    & = 
    \frac{v\kappa_{SH}}{2v_S^2 m_s^2 (4\pi)^2}\left(3 m_s^2+\frac{3 v^2\kappa_{SH}}{2v_S}-v_S \kappa_S\right)\llog m_s^2- \frac{\delta_{\rm CT}^{(1)}Z_{hs}}{2}\,.
\end{align}
\end{subequations}

\paragraph{\underline{Renormalisation in the OS scheme:}} The OS one- and two-loop counterterms for the SSM are defined in \cref{sec:SSM}. Their algebraic structure is rather involved and not particularly illuminating. Therefore, we only list analytic results required for the LSZ contributions, analogous to the \MS scheme above.
The relevant renormalised one-loop trilinear couplings read
\begin{subequations}
\begin{align}
    \left.\hat\lambda^{(1)}_{hhh}\right|^{\text{OS}} & = \frac{1}{(4\pi)^2}\frac{\kappa_{SH}^3v^3}{2 v_S^3 m_s^2} + \mathcal{O}(m_h^2/m_s^2)\,, \label{eq:SSM:oneloopresult}\\
    \left.\hat\lambda^{(1)}_{hhs}\right|^{\text{OS}} &=  -\frac{1}{(4\pi)^2}\frac{\kappa_{SH}^2v^2}{2 v_S^3 m_s^2}\left(
    3m_s^2 + \frac{3\kappa_{SH}v^2}{2v_S} - v_S\kappa_S
    \right) + \mathcal{O}(m_h^2/m_s^2)\,,
\end{align}
\end{subequations}
whereas the finite LSZ factors are rather simple,
\begin{subequations}
    \begin{align}
    \left.Z_{hh}^{(1)}\right|^{\text{OS}} & =
    \left.Z_{hh}^{(1)}\right|^{\MS}\,,\\
    \left.Z_{hs}^{(1)}\right|^{\text{OS}} & = \frac{\delta_{\rm CT}^{(1)}Z_{hs}}{2}\,,
\end{align}
\end{subequations}
due to the on-shell renormalisation of $\alpha$ which removes the $h-s$ mixing. The algebraic $\alpha$ counterterm, expanded to $\order{\epsilon^1}$, is given by
\begin{align}
    \delta_{\rm CT}^{(1)}\alpha^{\text{OS}} &= -\frac{\Sigma_{hs}(0)}{m_s^2} \nn \\
    & = -\frac{v\kappa_{SH}}{2v_S^2 m_s^2 (4\pi)^2}\left(3 m_s^2+\frac{3 v^2\kappa_{SH}}{2v_S}- v_S\kappa_S\right)\left(\llog m_s^2 - \frac{1}{\epsilon} + \epsilon\, (1-\llog m_s^2) \right) \,,
\end{align}
where the finite part coincides with the off-diagonal LSZ factor in the \MS scheme, $$Z_{hs}^{(1)}\big|^{\MS}=-\delta_{\rm CT}^{(1)}\alpha^{\OS}\,.$$

\paragraph{\textbf{\underline{Validation:}}} To further check our calculational framework for consistency, we computed all renormalised scalar one-loop one-, two-, three-, and four-point functions of the SSM in both, the \MS and the \OS scheme and found UV-finite results. However, computing quantities at the $n$-loop level that implicitly involve $\mu_S$ at the $(n-1)$-loop level, such as \eg $\hat{\Sigma}_{ss}^{(1)}$ or $\hat{\lambda}_{ssss}^{(1)}$, must be taken with care in the mixed OS-\MS scheme since one has to distinguish between the OS counterterm for the singlet tadpole coefficient, 
\begin{equation}
t_s=\left.\frac{\partial V_{\rm SSM}}{\partial s}\right|_{s=h=0}\,,
\end{equation}
and the counterterm of the linear operator $\tilde{t}_S \cdot S$ in \cref{eq:SSM:pot}. While $\delta_{\rm CT}^{(1)}t_s$ appears \eg in the renormalisation of $t_h$, \cf \cref{eq:SSM:dtadh2L}, the counterterm of $\mu_S$ is related to $t_s-\tilde{t}_S$ via the tree-level tadpole condition:
\begin{equation}
\label{eq:SSM:muS}
    \mu_S = \frac{t_s -\tilde{t}_S}{v_S} -
   v_S (\kappa_S + 2 \lambda_S v_S) - \frac{v^2}{2v_S}(\kappa_{SH} + \lambda_{SH} v_S)\,.
\end{equation}
Thus, $t_s$ and $\tilde{t}_S$ do not always appear in the same linear combination and need to be renormalised separately.
Considering for example the UV-divergent part of the one-loop self-energy $\Sigma_{SS}$ in the gauge basis, we can compute a counterterm for $\mu_S$ and indeed find full agreement with the \MS counterterm 
\begin{equation}
    (4\pi)^2\delta_{\rm CT}^{(1)}(\mu^2)^\MS=(2\kappa_S^2+6\mu_S^2\lambda_S)/\epsilon
\end{equation}
extracted from the RGEs for $\mu_S$ computed with \texttt{SARAH} within our approximation (taking care of scale-less integrals as discussed above). Applying a counterterm transformation onto the right-hand side of \cref{eq:SSM:muS} and using the \MS counterterms in \cref{eq:app:SSM:MScts} as well as renormalising $t_s$ in the OS scheme, \ie
\begin{align}
    \delta^{(1)}_{\rm CT}t_s^{\OS}=-t_s^{(1)}=(\kappa_S+6\,v_S\lambda_S)\textbf{A}_0(m_s^2)\,,
\end{align}
we can solve for $\delta_{\rm CT}^{(1)}\tilde{t}_S$ and find
\begin{equation}
\label{eq:SSM:dtadSprime}
    (4\pi)^2\delta_{\rm CT}^{(1)}\tilde{t}_S = \frac{\kappa_S}{2\epsilon}\left(v_S\, \kappa_S+\frac{v^2\kappa_{SH}}{2}+m_s^2\right)\,,
\end{equation}
which is non-zero for $\kappa_S\neq0$. Likewise, the linear combination $\delta_{\rm CT}^{(1)}\tilde{t}_S-\delta_{\rm CT}^{(1)}t_s$ enters the renormalisation of the singlet self-energy $\Sigma_{ss}$ and its cubic and quartic self-couplings, which are all rendered UV-finite by \cref{eq:SSM:dtadSprime}. It should be stressed that the choice of renormalising $\tilde{t}_S$ is not unique. It is related via a shift symmetry to the parameters $v_S$ and $\kappa_S$ which could equally be used to obtain UV-finite results.

\section{\texorpdfstring{\texttt{Tintegrals Python}}{Tintegrals Python} package}
\label{app:Tint}

\Tint is a \py package to evaluate the $T$ integrals defined in \cref{eq:Tint} numerically for vanishing external momenta. The \Tint source code is hosted at
\begin{center}
  \url{https://gitlab.com/anybsm/tint}.
\end{center}
The code is most easily used by installing the corresponding \py package by running
\begin{minted}[bgcolor=bg]{bash}
pip install Tintegrals
\end{minted}
which will automatically download and install \Tint as well as the necessary dependencies. Running the code requires at least \py version 3.5.

The \Tint module is loaded via \eg
\begin{minted}[bgcolor=bg]{python}
import Tintegrals as Tint
\end{minted}
The module includes functions to numerically evaluate the UV-finite parts of all $T$ integrals needed to evaluate the generic expressions derived in this work. This implementation includes all the special cases derived during the integral reduction (see the discussion in \cref{sec:integral_reduction}). To decide whether two masses, $m_1$ and $m_2$, should be considered numerically equal, the module checks whether
\begin{align}
    |m_1 - m_2| < \rm{max}(\epsilon_\text{rel}(m_1 + m_2), \epsilon_{\rm{diff}})\,.
\end{align}
where by default, $\epsilon_\text{rel} = 10^{-10}$ and $\epsilon_\text{diff} = 10^{-12}$.

Using this criterion, the loop integrals are then evaluated \eg via
\begin{minted}[bgcolor=bg]{python}
Tint.setMudim(125**2)
Tint.T134fin(125, 125, 125)
\end{minted}
where in the first step the renormalisation scale is set to $125\gev$. In the second step, the value of the $T_{134}(m_h,m_h,m_h)$ scalar integral for an internal mass of $125\gev$ is calculated.

Comprehensive documentation outlining the complete range of functions is accessible at
\begin{center}
\url{https://anybsm.gitlab.io/tint}.
\end{center}


\clearpage
\printbibliography

\end{document}